\documentclass[twocolumn,showpacs,preprintnumbers,nofootinbib,superscriptaddress]{revtex4}
\usepackage{epsfig,amsmath,amssymb}
\usepackage{graphicx}
\usepackage{dcolumn}
\usepackage{bm}
\usepackage{subfigure}
\usepackage{color}
\usepackage{multirow}
\usepackage{textcomp}
\newcommand{\be}{\begin{equation}}    
\newcommand{\ee}{\end{equation}}
\newcommand{\beq}{\begin{eqnarray}}
\newcommand{\eeq}{\end{eqnarray}}
\newcommand{\beqn}{\begin{eqnarray*}}
\newcommand{\eeqn}{\end{eqnarray*}}
\newcommand{\msun}{M_{\odot}}
\newcommand{\mnras}{Monthly Notices of the RAS}
\newcommand{\apjl}{Astrophysical Journal, Letters}

\newcommand{\MWEG}{$\rm{MWEG}^{-1}\times \rm{Myr}^{-1}$ }

\newcommand{\Mpc}{$\rm{Mpc}^{-3}\times \rm{Myr}^{-1}$ }
\newcommand{\mmm}{\rm{Mpc}^{-3}\times \rm{Myr}^{-1} }

\newcommand{\e}{$\eta$ }

\definecolor{magenta}{cmyk}{0.1,0.8,0,0.1} 

\begin{document}

\title{Compact binaries detection rates from gravitational wave
interferometers: comparison of different procedures}
\author{G. Corvino}
\email{giovanni.corvino@roma1.infn.it}
\author{V. Ferrari}
\author{S. Marassi}
\affiliation{Dipartimento di Fisica `G. Marconi', Sapienza Universit\`a di Roma and Sezione INFN Roma1, Piazzale Aldo Moro 5, Roma, 00185, Italy}
\author{R. Schneider}
\affiliation{INAF, Osservatorio Astronomico di Roma, via di Frascati 33, 00040 Monteporzio Catone, Italy}

\date{\today}
\label{firstpage}

\begin{abstract}
In this paper we perform a detailed analysis of the effect of various
approximations which have been used in the literature to compute the
detection rates of compact binary coalescences 
for interferometric gravitational wave detectors.  
We evaluate the detection rates for the coalescence of BH-BH,
NS-NS, and BH-NS binaries taking into account their specific statistical
properties obtained from population synthesis models,
the cosmic star formation rate history, and the
effects of redshift on the emitted gravitational wave signals. 
The results are compared  with those obtained with 
procedures that are based on different levels of approximations, such as adopting averaged
values for the total mass and symmetric mass ratio for all the systems
of a binary population, using these to compute the horizon distance for
individual detectors, or estimating the coalescence rate density within
this distance by its local value. 
We show that these approximations introduce a bias which depends not 
only on the approximation which is used, but also on the statistical 
properties of the considered binary population.
\end{abstract} \pacs{to be inserted} \maketitle

\section{Introduction}
\label{sec:intro}
The second generation of gravitational wave interferometers, Advanced 
Virgo\footnote{https://wwwcascina.virgo.infn.it/advirgo/} and 
LIGO\footnote{http://www.ego-gw.it/, http://www.ligo.caltech.edu}, 
will push ten times farther their observational reach, opening a much
wider region of the Universe to observations. Furthermore,
the third-generation  ground-based observatory Einstein Telescope
(ET)\footnote{ET: www.et-gw.eu},  which is currently in a design study
phase, is expected to improve the
sensitivity by a further factor of ten \cite{PunAbeAce2010,RegDenDel2012}.
This will allow to detect gravitational wave (GW) signals
emitted by astrophysical sources, for instance the coalescence of compact
binaries in which we are primarily interested, 
with  sufficiently high signal-to-noise ratios
(SNR) and rates, to allow precise astronomical studies of these 
sources \cite{SatAbeAce2012}.
In addition, the high sensitivity of these  detectors 
will allow to see beyond our local universe,
accessing regions of redshift  where the role of the 
star formation history and of cosmological models becomes important,
and could be tested (\cite{SatSchVan2010,TayGai2012,TayGaiMan2012} 
and references therein).

The detection of a GW signal is based on the correlation of the 
output data stream of the detector with an expected signal, 
which is typically in the form of a template. 
Given the high number of parameters needed to
model a source, a complete survey of the parameter space is often
impossible. For this reason it is important to know
what are the systems that are more likely to be detected and
their expected properties \cite{AbaAbbAbb2011,AbaAbbAbb2012b}.
In recent years several papers have been published where  
the coalescence rates of compact binaries,
and  the corresponding detection rates by GW detectors, 
have been estimated by using several approximations
\cite{AbaAbbAbb2010b,GaiManMill2011,MarSchCor2011,ManOsh2010,BulBelRud2004,
BulGonBel2004,VosTau2003,OshKalBel2010,OshBelKal2008,OshKimKal2008,
BelBulBai2011,BelDomBul2010,BelTaaKal2007,OshKimFra2005}.

In this paper we study in detail the effects of these approximations on
the detection rates of advanced detectors,
also showing to what extent these effects depend on 
the binary  population model which is used. 
We choose, as an example, two  models generated 
by the code {\tt Startrack}, and publicly 
available\footnote{http://www.syntheticuniverse.org}.
Details on the physical inputs of the simulation  are given in
\cite{DomBelFry2012}.
The binary populations  are the result of the evolution of
a large sample ($N = 2\cdot10^{6}$) of ``zero-age'' binaries.

Using these data, we compute the coalescence 
and detection rates of binaries composed of two
black holes, BH-BH, two neutron stars, NS-NS, a
black hole and a neutron star, BH-NS, using a 
completely consistent procedure based on the appropriate use of 
the statistical properties of the populations, and on the redshift 
dependence of the star formation rate history 
\cite{FlaHug1998,SchFerMat2001,MarSchCor2011,GaiManMill2011}. 
The detection rates are given for Advanced 
Virgo and LIGO, and for ET.

These rates are then compared with those computed
by applying, to the same populations, the
different approximations used in the literature,
to evaluate the bias they introduce
(if any), to show how this varies for the Advanced
detectors in different configurations, 
and to understand its dependence on the population  properties.

The statistical properties of coalescing binary populations 
depend on the physical inputs of the population synthesis codes, which
may vary significantly from one model to another. The choice of the two
models we consider in this paper, together with their main statistical 
properties, are briefly described in section \ref{sec:ratecomp}. 

We will show that in order to obtain detection rate estimates which
depend only on the properties of the binary population,
we need to use a fully  consistent approach, since each of the
approximations used in the literature leads to over/underestimate 
these rates by amounts which depend on the population properties.

The plan of the paper is the following. 
In section \ref{sec:ratecomp} we describe the fully consistent 
procedure  to compute the 
coalescence and detection rates for a simulated population of compact
binaries.
In section \ref{sec:comparison} we describe various approximations
 used to evaluate the rates
\cite{AbaAbbAbb2010b,BelTaaKal2007,BelDomBul2010}, apply them to the two
population models we  consider, and compare the results with those found with no
approximation. 
Concluding remarks are given in section \ref{sec:conclusions}.

\section{Rate computation procedure}
\label{sec:ratecomp}
The binary population models we use to compute coalescence and
detection rates, are chosen from the Standard model of
\cite{DomBelFry2012},
since this model is the fiducial one, being based on conservative assumptions. 
We consider the two submodels A which correspond to progenitor stars with solar
and subsolar metallicity. The corresponding
populations exhibit different mass distributions, 
and this is the feature which is of major interest for us.
Hereafter we will refer to these two models as model A (solar
metallicity) and model B (subsolar metallicity).

Here we report some statistical properties of 
each model, which will be useful in the following to understand the
results. First of all,
it should be  mentioned that, as a result of different physical inputs
(which we do not discuss here), the number of 
binaries coalescing within the Hubble time, differs in the two models:
for model A there are
1753 BH-BH binaries, 5084 NS-NS and 345 BH-NS;
for model B the population is made of
15650 BH-BH, 1767 NS-NS and 769 BH-NS binaries. 
Thus, the lower metallicity used in model B causes an
increase of the number of BHs that are formed and of their mean mass as well.

The average total mass $\bar{M}$ and symmetric
mass ratio $\bar{\eta}$ are: \\
for model A 
\begin{eqnarray}
\label{mod1}
&\bar{M}=15.41 \msun&  \bar{\eta}=0.248\quad\hbox{for~BH-BH}
\\\nonumber
&\bar{M}=2.43 \msun&  \bar{\eta}=0.248\quad\hbox{for~NS-NS}
\\\nonumber
&\bar{M}=9.94 \msun&  \bar{\eta}=0.150\quad\hbox{for~BH-NS}
\end{eqnarray}
for model B
\begin{eqnarray}
\label{mod2}
&\bar{M}=30.66 \msun&  \bar{\eta}=0.246\quad\hbox{for~BH-BH}
\\\nonumber
&\bar{M}=2.50 \msun&  \bar{\eta}=0.249\quad\hbox{for~NS-NS}
\\\nonumber
&\bar{M}=11.72 \msun&  \bar{\eta}=0.117\quad\hbox{for~BH-NS}
\end{eqnarray}

The procedure we adopt to compute the detection rates from the
population synthesis statistics is the following.
Starting from the star formation rate density at a given redshift,
$\dot{\rho}_{\star}(z)$ (in units of $M_\odot$
yr$^{-1}$Mpc$^{-3}$), we derive the binary progenitors birth rate per
comoving volume (in units of yr$^{-1}$ Mpc$^{-3}$) as, 
\be
\frac{dR}{dtdV}(z)= A\times \dot{\rho}_{\star}(z) \label{birth} 
\ee
\noindent 
where 
\be 
A=\frac{f_{\rm bin}\times f_{\rm sim}}{2\langle
m_\star \rangle}~, \label{costbirth} 
\ee 
\noindent 
$f_{\rm bin}$ is the
binarity fraction, $\langle m_\star \rangle$ is
the average stellar mass and $f_{\rm sim}$ is the fraction of binaries
simulated by the population synthesis code.  The latter quantity
accounts for the fact that the range of progenitor masses used in the 
simulation is different from the range of validity of the adopted IMF.
The factors $f_{\rm sim}$ and
$\langle m_\star \rangle$ depend on the adopted stellar Initial Mass
Function (IMF), $\Phi(m)$, as
\be f_{\rm sim}=\frac{\int^{100}_{8} dm \Phi(m)}{\int^{100}_{0.1} dm
\Phi(m)}, \ee
\be \langle m_{\star} \rangle=\frac{\int^{100}_{0.1}dm m
\Phi(m)}{\int^{100}_{0.1}dm \Phi(m)}~.  \ee \noindent In \cite{DomBelFry2012}
the authors set $f_{\rm bin}=2/3$ and initialize the stellar progenitors in 
the mass range $[5-150]\msun$ according to a Kroupa IMF.

We extract from the simulated sample of $N_{tot}$ binaries, those
systems with total mass $\in (M,M+dM)$, symmetric mass ratio $\in
(\eta,\eta+d\eta)$, and delay time $\in(t_d,t_d+dt_d)$, which form BH-BH,
NS-NS, and BH-NS systems; the delay time is defined as the time
interval between the formation of the binary progenitor system and its
coalescence. Their number is $N_{i}(M,\eta,t_d)$, where $i$ refers to
BH-BH, NS-NS, BH-NS systems.

The number of binaries coalescing at redshift $z_{c}$, with total mass
$M$ and symmetric mass ratio \e, per unit of $M$, \e and comoving volume
is ($\rm{Mpc}^{-3} \rm{yr}^{-1} \rm{M}_{\odot}^{-1} \eta^{-1}$):

\begin{equation}
\dot{r}_{\rm{coal}}(M,\eta,z_{c}) = \int  
\frac{A\times \dot{\rho}_{\star}(z_f)}{1+z_f}
\frac{N_i(M,\eta,t_d)}{N_{tot}}~dt_d~;
\label{rcoal_def} 
\end{equation}
\noindent
the redshifts of progenitor binary formation, $z_f$, and of
binary coalescence, $z_c$, are related by the delay time as follows

\begin{equation} t_d =
\int_{z_{c}}^{z_{f}}\left|\frac{dt}{dz}\right|dz~, 
\end{equation}
\noindent
and
\begin{equation}
 \left|\frac{dt}{dz}\right| =
\frac{1}{H_{0}(1+z)\sqrt{\Omega_{m}(1+z)^{3}+(1-\Omega_{m})}}~.
\label{dtdz}
\end{equation}
\noindent
In this work we assume $H_{0}=73 \;\rm{km}^{-1}\,\rm{Mpc}^{-1}$ and
$\Omega_{m}=0.24$, consistently with the choice done in \cite{TorFerSch2007} 
to derive our adopted cosmic star formation history, which reproduces 
the observational data available at $z< 8$\cite{BouIllOes2010}.

The coalescence rate density $\dot{\rho}_c(z_c)$ 
($\rm{Mpc}^{-3}\rm{yr}^{-1}$), i.e. the number of compact binaries
coalescing at redshift $z_c$, per unit time and unit comoving volume, is
\beq
\label{coalratedens}
 \dot{\rho}_c(z_c) &=& \iint \dot{r}_{\rm{coal}}(M,\eta,z_c)dMd\eta
\\\nonumber
&=&A\int dM\int d\eta \int \frac{ \dot{\rho}_{\star}(z_f)}{1+z_f}
\frac{N_i(M,\eta,t_d)}{N_{tot}}~dt_d ~.
\eeq
\noindent
\begin{figure}[h]
 \centering
 \includegraphics[width=8.5cm,height=6cm]{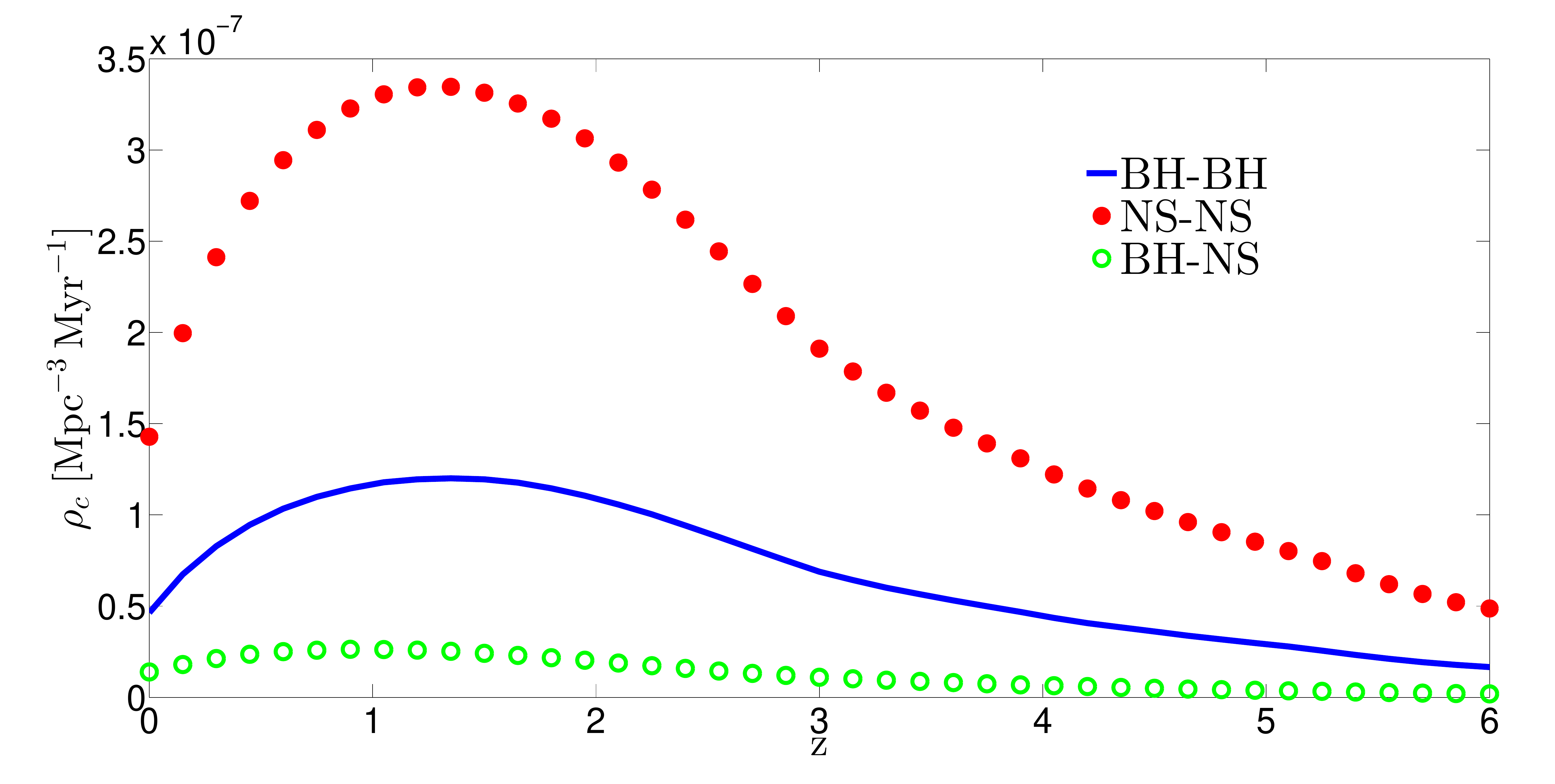}
 \includegraphics[width=8.5cm,height=6cm]{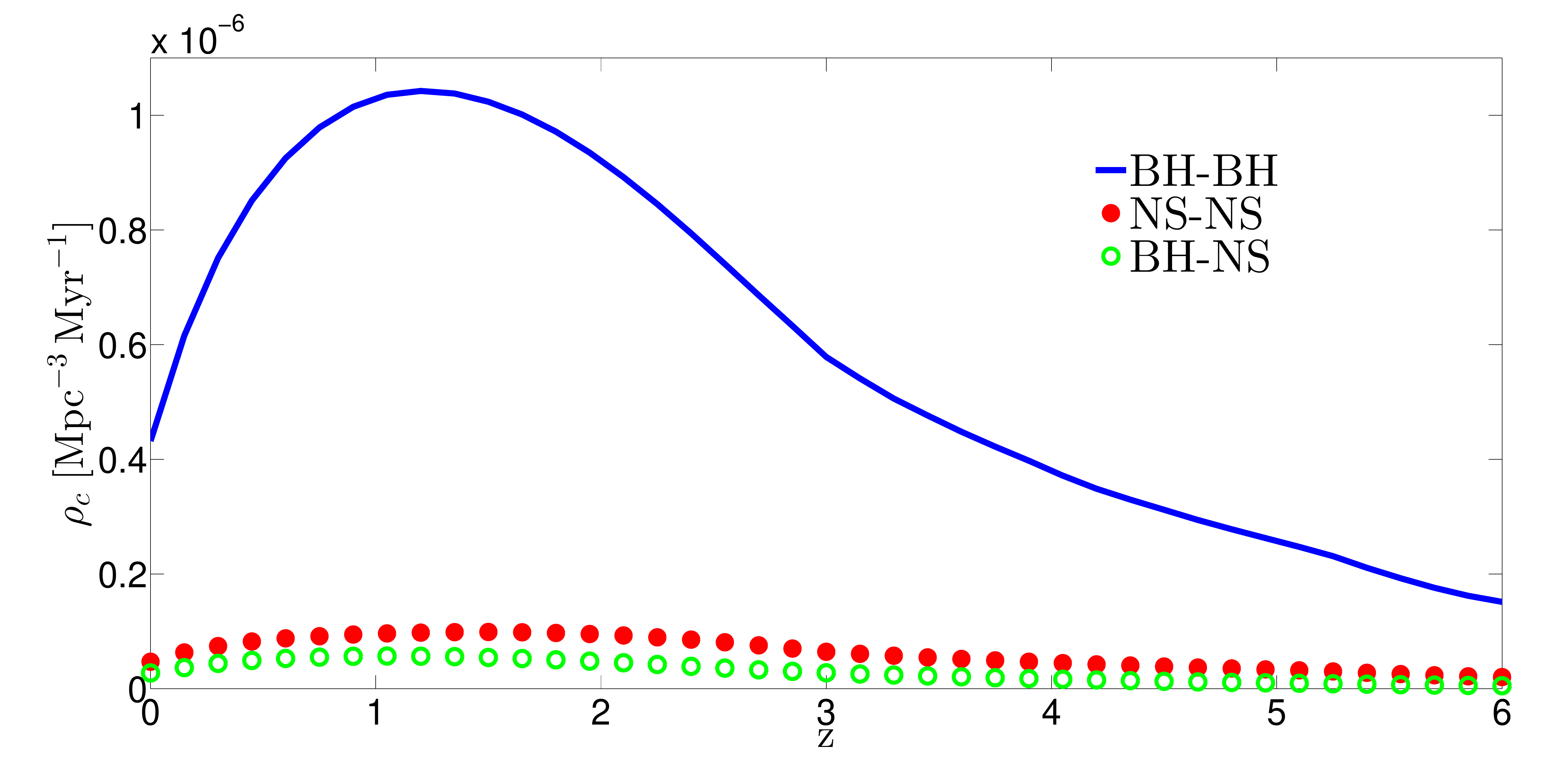}
 \caption{(color online) The coalescence rate density
computed with Eq.~(\ref{coalratedens}), is plotted
versus redshift,  respectively  for BH-BH,  NS-NS  and BH-NS binaries,
for model A and B.
}
\label{FIG1}
\end{figure}
In Fig.~\ref{FIG1} we plot $\dot{\rho}_c$ as a function of
redshift for model A (upper panel) and model B (lower panel). We define the value
$\dot{\rho}^{\rm{local}}_c=\dot{\rho}_c(z_c=0)$ as the ``local merger
rate per unit comoving volume''.  For model A we find 
\beq
\label{rholocal}
&\dot{\rho}^{\rm{local}}_c=4.64\times 10^{-2}~\mmm &\hbox{(BH-BH)}
\\\nonumber
&\dot{\rho}^{\rm{local}}_c=1.43\times 10^{-1}~\mmm &\hbox{(NS-NS)}
\\\nonumber
&\dot{\rho}^{\rm{local}}_c=1.40\times 10^{-2}~\mmm &\hbox{(BH-NS)}.
\eeq
and for model  B
\beq
\label{rholocal2}
&\dot{\rho}^{\rm{local}}_c=4.32\times 10^{-1}~\mmm &\hbox{(BH-BH)}
\\\nonumber
&\dot{\rho}^{\rm{local}}_c=4.69\times 10^{-2}~\mmm &\hbox{(NS-NS)}
\\\nonumber
&\dot{\rho}^{\rm{local}}_c=2.73\times 10^{-2}~\mmm &\hbox{(BH-NS)}.
\eeq
To compute the detection rate, we sum over the sources which are
within a given interferometer reach. 
Assuming that the signals are extracted from noise using the standard
matched filtering technique, the Signal to Noise Ratio (SNR), averaged
over sky positions and angles is \cite{FlaHug1998}:

\begin{equation} 
\langle SNR\rangle^2 =  \frac{3}{2}\left\{
\frac{2}{5}\frac{(1+z)^2}{\pi^2 d_L(z)^2}
\int_{0}^{\infty} df \frac{1}{f^2S_{h}(f)} \frac{dE_e}{df_e}
[(1+z)f]
\right\}~,
 \label{SNRdef}
\end{equation}
\noindent
where $f$ is the observed frequency, related to the source emission
frequency $f_e$ by
\begin{equation}
f=\frac{f_e}{1+z}~,
\label{redf}
\end{equation}
$dE_e/df_e$ is the source energy spectrum,
and $S_h$ is the detector noise spectral density.
\noindent
For BH-BH systems we shall use  the most recent
hybrid  waveforms produced by numerical relativity groups
 \cite{AjiHanHus2011}. These describe
the full inspiral-merger-ringdown  phases of coalescence, accounting
also for non-precessing spins. However, since the simulated populations
we use do not include spins, we shall assume that all BH-BH systems have
zero spin.
For NS-NS and BH-NS binaries we use the chirp
waveform, truncated to the ISCO.
\begin{table*}[ht]
  \begin{center}
    \begin{tabular}{|c|c|c|c|c|c|c|}
      \hline
      \multicolumn{7}{|c|}{\multirow{2}*{No Approximation detection
rates $N_{\rm{det}}$ $[\rm{yr}^{-1}]$}}\\
      \multicolumn{7}{|c|}{} \\
      \hline
      \multirow{3}*{detector}   &
\multicolumn{2}{|c|}{\multirow{2}*{BH-BH}} & 
\multicolumn{2}{|c|}{\multirow{2}*{NS-NS}} &
\multicolumn{2}{|c|}{ \multirow{2}*{BH-NS}} \\
&\multicolumn{2}{|c|}{} & 
\multicolumn{2}{|c|}{} &
\multicolumn{2}{|c|}{} \\
      \cline{2-7}
&model A& model B &model A& model B&model A& model B\\
      \hline
	AVirgo & 73 & $ 5.67\cdot 10^{3}$  & 2.3  & 0.82 & 3.7  & 6.61 \\
      \hline
	ALIGO1 & 203 & $ 1.52\cdot 10^{4}$   & 6.0  &2.10 & 9.2  &  17.8\\
      \hline
	ALIGO2 & 134  &$ 1.07\cdot 10^{4}$   & 3.9  &1.35 & 6.2  & 12.0 \\
      \hline
	ALIGO3 & 105   &$ 8.34\cdot 10^{3}$  & 3.1  &1.08 & 5.0  & 9.60 \\
      \hline
	ALIGO4 	& 254   &$ 1.84\cdot 10^{4}$  & 7.6  &2.66 & 11.8&  22.0 \\
      \hline
ET\_B 	& $1.80\cdot 10^{5}$&$ 1.56\cdot 10^{6}$  &$ 2.65\cdot 10^{4}$ & 
$8.01\cdot 10^{3}$& 
$1.79\cdot 10^{4}$  & \ $3.62\cdot 10^{4}$\\
      \hline
ET\_C  	& $1.84\cdot 10^{5}$&$ 1.59\cdot 10^{6}$  &$ 4.48\cdot 10^{4}$ &
$1.35\cdot 10^{4}$&
$2.47\cdot 10^{4}$  & \ $5.39\cdot 10^{4}$\\
      \hline
ET\_D	&$ 1.81\cdot 10^{5}$&$ 1.57\cdot 10^{6}$  &$ 2.64\cdot 10^{4} $&
$7.95\cdot 10^{3}$& 
$1.87\cdot 10^{4}  $& $3.86\cdot 10^{4}$\\
      \hline
    \end{tabular}
    \caption{Detection rates computed from Eq.~(\ref{Napprox0}), 
for the simulated populations of BH-BH, NS-NS  and
BH-NS (model A and B) and various detector sensitivities 
(first column) (see text).}
\label{table1}
\end{center} 
\end{table*}
Moreover we will consider various noise curves of advanced detectors,
namely
Advanced LIGO (ALIGO) in four
of its planned configurations \cite{Harry2010}\footnote{see LIGO
document LIGO-T0900288-v3}, namely  zero detuning high
laser power (ALIGO1), optimized for 30-30 solar mass bh binary
(ALIGO2), no signal recycling mirror (ALIGO3),
and optimized for NS-NS binaries (ALIGO4),
Advanced Virgo\footnote{wwwcascina.virgo.infn.it/
advirgo/docs/AdV\_refsens\_100512.txt}(AVirgo), and
ET\footnote{http://www.et-gw.eu/etsensitivities} in three different
configurations namely ET\_B, ET\_C and ET\_D.

We define $ z_{\rm max} (M,\eta)$ as the maximum redshift at which a
system with assigned $(M,\eta)$ is detectable with SNR=8, averaging over
position and orientation;  $z_{\rm max} (M,\eta)$ is obtained by inverting
Eq.~(\ref{SNRdef}).  $ D_{\rm max} (M,\eta)$ is 
the corresponding luminosity distance.

Moreover, we define  $z_{\rm h}(M,\eta)$ and $D_{\rm h}(M,\eta)$,
as the horizon redshift and horizon distance, computed as before, but
assuming optimal orientation.

The detection rate per unit total mass $M$ and
mass ratio $\eta$, $N_{M,\eta}$, expressed in units 
$[\rm{yr}^{-1} \rm{M}_{\odot}^{-1} \eta^{-1}]$, is found by integrating 
the function $\dot{r}_{\rm{coal}}(M,\eta,z_c)$ up to 
$z_{\rm max}(M,\eta)$, as follows:

\begin{equation} 
N_{M,\eta}(M,\eta) =
\int_{0}^{ z_{\rm max}(M,\eta)}\dot{r}_{\rm{coal}}(M,\eta,z_c)\frac{dV}{dz_c}dz_c~,
\label{Metarate} 
\end{equation} 
\noindent
where the function
\begin{equation}
 \frac{dV}{dz} = 4\pi c \frac{d_{L}(z)^{2}}{1+z}\left|\frac{dt}{dz}\right| \label{covol}
\end{equation}
\noindent 
is the comoving volume element and $d_{L}(z)$ is the luminosity distance.
It has to be noted that 
$N_{M,\eta}$ depends on $M$ and \e, not only through the 
coalescence rate $\dot{r}_{\rm{coal}}$, but also through the integration limit. 
This means that, for each system (i.e. for each value of $M$ and \e), the 
coalescence rate is integrated up to a different distance. 

The integrated detection rate, i.e. the final output of
our procedure,  in units  [yr$^{-1}$] is:
\begin{eqnarray}
 N_{\rm{det}} &=& \int dM \int  N_{M,\eta}(M,\eta) d\eta \label{Napprox0}\\
\nonumber
  &=& \int dM \int d\eta
\int_{0}^{z_{\rm max} (M,\eta)}\dot{r}_{\rm{coal}}(M,\eta,z_c)\frac{dV}{dz_c}dz_c~.
\end{eqnarray} 
In the following we will refer to the results obtained with this procedure as 
the ``no-approx'' rates.

In Table~\ref{table1}, we show the detection rates obtained for
the three considered families of compact binaries, using  model A and B. 
The uncertainty in these results, due to numerical integration, is of a few percent. 

The data show that, for model A and B,
the highest rates are attained by BH-BH systems, 
both for advanced and for third generation detectors.

As already noted in \cite{BelDomBul2010}, where detection
rates where computed only for advanced detectors,
including metallicity effects in stellar evolution results in a much higher
detection chance for BH-BH and BH-NS coalescing binaries.
This is because  in model B the number of these systems is larger than in
model A, and the average mass of systems including
BHs is higher (see Eqns.~(\ref{mod1}) and (\ref{mod2}) and the text above).

Finally, in Table~\ref{table2} we give the values of the
horizon distance, $\bar{D}_{\rm h}\equiv D_{\rm h}(\bar{M},\bar{\eta})$,
which  corresponds to the mean values of mass and mass
ratio  given in Eqns.~(\ref{mod1})  and (\ref{mod2}); this quantity
will  be useful in the following discussion. 

\begin{table}[t]
  \begin{center}
    \begin{tabular}{|c|c|c|c|c|c|c|}
      \hline
      \multicolumn{7}{|c|}{\multirow{2}*{Horizon distances $\bar{D}_{\rm
h}$ [Mpc]}}\\
      \multicolumn{7}{|c|}{} \\
      \hline
\multirow{2}*{detector}				& \multicolumn{3}{|c|}{model A}
& \multicolumn{3}{|c|}{model B} \\
      \cline{2-7}
         &BH-BH& NS-NS&BH-NS &BH-BH& NS-NS&BH-NS\\
      \hline
      AVirgo   	 & 1980 &  371 &  959  & 4030 & 381 & 955\\
      ALIGO1  	 & 2980 &  518 &  1390 & 6110 & 533 & 1390\\
      ALIGO2  	 & 2520 &  445 &  1200 & 5210 & 458 & 1200 \\
      ALIGO3   	 & 2290 &  412 &  1100 & 4580 & 423 & 1100 \\
      ALIGO4   	 & 3260 &  562 &  1510 & 6770 & 578 & 1500\\
      \hline
    \end{tabular}
    \caption{Horizon distances of advanced detectors (in Mpc), computed
for the mean values of mass and mass ratio  given in
Eqns.~(\ref{mod1})  and (\ref{mod2}), for the three families of binaries 
and for models A and B.}
    \label{table2}
  \end{center}
\end{table}

\section{Comparison with other procedures}
\label{sec:comparison}
Various approaches have been used  in the literature to compute detection rates,
starting from the coalescence rates produced by population synthesis
simulations  \cite{AbaAbbAbb2010b,ManOsh2010,BulBelRud2004,
BulGonBel2004,VosTau2003,OshKalBel2010,OshBelKal2008,OshKimKal2008,
BelBulBai2011,BelDomBul2010,BelTaaKal2007,OshKimFra2005}.
A common feature of these approaches is that the detection rate is evaluated by
multiplying the merger rate 
by a properly defined  detection volume $V_{\rm{h}}$, rather than using Eq.~(\ref{Napprox0}). 
\begin{table*}[ht]
  \begin{center}
    \begin{tabular}{|c|c|c|c|c|c|c|c|c|c|c|c|c|}
      \hline
      \multicolumn{13}{|c|}{\multirow{2}*{Approximation 1: detection
rates $[\rm{yr}^{-1}]$}}\\
      \multicolumn{13}{|c|}{} \\
      \hline
& \multicolumn{6}{|c|}{model A}&\multicolumn{6}{|c|}{model B}\\
      \cline{2-13}
      \multirow{2}*{detector}   & \multicolumn{2}{|c|}{BH-BH}
& \multicolumn{2}{|c|}{NS-NS}           & \multicolumn{2}{|c|}{BH-NS}
&\multicolumn{2}{|c|}{BH-BH}
& \multicolumn{2}{|c|}{NS-NS}           & \multicolumn{2}{|c|}{BH-NS}
 \\
      \cline{2-13}
                                & \multirow{2}*{$N_{\rm{det}}^{1}$} &
\multirow{2}*{$N_{\rm{det}}$} & \multirow{2}*{$N_{\rm{det}}^{1}$} &
\multirow{2}*{$N_{\rm{det}}$} &
                                  \multirow{2}*{$N_{\rm{det}}^{1}$} &
\multirow{2}*{$N_{\rm{det}}$} 
& \multirow{2}*{$N_{\rm{det}}^{1}$} &
\multirow{2}*{$N_{\rm{det}}$} & \multirow{2}*{$N_{\rm{det}}^{1}$} &
\multirow{2}*{$N_{\rm{det}}$} &
                                  \multirow{2}*{$N_{\rm{det}}^{1}$} &
\multirow{2}*{$N_{\rm{det}}$} 
\\
      & & & & & &  & & & & & &\\
      \hline
	AVirgo  & 79   & 73   & 2.4& 2.3  & 3.8  & 3.7
&$3.98\cdot10^3$& $5.67\cdot10^3$ & 0.78 & 0.82 & 7.86 & 6.61\\	
      \hline
	ALIGO1 	& 210  & 203  & 6.2  & 6.0  & 9.3 & 9.2
&$1.07\cdot10^4$& $1.52\cdot10^4$ & 2.01 & 2.10 & 21.0 & 17.8\\	
      \hline
	ALIGO2	& 140  & 134  & 4.0  & 3.9  & 6.5  & 6.2
&$7.17\cdot10^3$& $1.07\cdot10^4$ & 1.30 & 1.35 & 14.1 &12.0\\	
      \hline
	ALIGO3  & 107  & 105  & 3.2  & 3.1  & 5.3  &5.0
&$5.36\cdot10^3$&$8.34\cdot10^3$     & 1.04 & 1.08& 11.3 & 9.60 \\	
      \hline
	ALIGO4 	& 260  & 254  & 7.9  & 7.6  & 12.3 &11.8
&$1.36\cdot10^4$& $1.84\cdot10^4$    & 2.55  & 2.66 & 25.2 & 22.0\\	
      \hline
    \end{tabular}
\caption{ The detection rate $N_{\rm{det}}^{1}$, evaluated from
Eq.~(\ref{det2}) (Approx 1) for model A and B.
$N_{\rm{det}}^{1}$ for BH-BH systems (column 2), is
compared to the no-approx rates, $N_{\rm{det}}$, computed from
Eq.~(\ref{Napprox0}) (column 3).  The additional columns refer to the
same quantities evaluated for NS-NS and BH-NS  binaries.}
\label{table3} 
\end{center} 
\end{table*}

In \cite{AbaAbbAbb2010b}
the authors adopt the following procedure.
Assuming that all black holes have $M=10~M_\odot$ and all
neutron stars have $M=1.4~M_\odot$, they define the average horizon distance
$D_{\rm h}$,
such that a binary system is detected with a $SNR\geq 8$ by a given
detector. 

A detection volume is defined as the Euclidean volume within 
$D_{\rm h}$:
\begin{equation}
 V_{\rm{h}} = \frac{4}{3}\pi
\left(\frac{D_{\rm h}}{\rm{Mpc}}\right)^{3}. \label{Vhor}
\end{equation}
\noindent
The detection rate is then computed by multiplying
the Galactic coalescence rate, $\dot{\rho}_c^{\rm{gal}}$ 
expressed in units Myr$^{-1}$ MWEG$^{-1}$ (Milky Way Equivalent
Galaxy), by the number of MWEGs inside $V_{\rm{h}}$, $G(D_{\rm h})$,
\begin{equation}
 N_{\rm{det}} = \dot{\rho}_c^{\rm{gal}} \times G(D_{\rm h}), \label{det1}
\end{equation}
\noindent
where
\begin{equation}
 G(D_{\rm h}) = 
V_h \times (0.0116)\times
\frac{1}{2.26^3}, \label{Ng}
\end{equation}
\noindent
The numerical factor $0.0116$ is an estimate of the local number density of
MWEGs in units $\rm{Mpc}^{-3}$ obtained in \cite{KopHanKal2008}; the
factor $1/2.26^3$ is needed to average over sky position and
orientation. This factor and the  distance $D_{\rm h}$ in
Eq.~(\ref{Vhor}), in ref. \cite{AbaAbbAbb2010b} 
are computed neglecting the effect of redshift (see their footnote 90). 

The Galactic coalescence rate, $\dot{\rho}_c^{\rm{Gal}}$ can be estimated
following two different procedures, i.e. by using a constant Galactic star formation 
rate \cite{NelYunPor2001a,OshKalBel2007}, or by normalizing the number of binaries 
produced by a population synthesis code to the observed Galactic supernova rate 
\cite{MarSchCor2011,BelTaaKal2007}. 
Using the former procedure, in \cite{DomBelFry2012} they find
\begin{itemize}
\item
$\dot{\rho}_c^{\rm{Gal}}=8.2$ \MWEG for BH-BH,
\item
$\dot{\rho}_c^{\rm{Gal}}=23.5$ \MWEG for NS-NS,
\item
$\dot{\rho}_c^{\rm{Gal}}=1.6$ \MWEG for BH-NS. 
\end{itemize}
for model A and
\begin{itemize}
\item
$\dot{\rho}_c^{\rm{Gal}}=73.3$ \MWEG for BH-BH,
\item
$\dot{\rho}_c^{\rm{Gal}}=8.1$ \MWEG for NS-NS,
\item
$\dot{\rho}_c^{\rm{Gal}}=3.4$ \MWEG for BH-NS. 
\end{itemize}
for model B.
If the detection rate is estimated using the local coalescence rate per unit 
volume, $\dot{\rho}_c^{\rm{local}}$, in units \Mpc, Eq.~(\ref{det1}) reads
\begin{equation}
 N_{\rm{det}} = \dot{\rho}_c^{\rm{local}} \times
V_{\rm{h}} \times \frac{1}{2.26^3}. \label{det2}
\end{equation}
\noindent
It should be noted that the ratio between the local and the Galactic
rates depends on the adopted cosmic star formation history model, on the
Galactic star formation rate, as well as on the statistical properties of
the binary population (delay time distribution). In
\cite{AbaAbbAbb2010b} a constant value of
$\dot{\rho}_c^{\rm{local}}/\dot{\rho}_c^{\rm{gal}}=0.0116$, has been adopted
for all compact binaries considered in their study.
In summary, the approximations done in \cite{AbaAbbAbb2010b} to estimate
the detection rates predicted by different binary populations and
different detectors are: 
\begin{enumerate} 
 \item  Approx 1: use of a single horizon
  distance $D_{\rm h}$ for all systems of a given binary population,
corresponding to the mean values of mass and mass 
ratio. 
 \item  Approx 2: evaluation of the detection rate
by multiplying  the local value of the coalescence rate density,
$\dot{\rho}_c(z=0)$, by the Euclidean volume, within the horizon distance
$D_{\rm h}$, as in Eqns.~(\ref{det1}) or
(\ref{det2}).
 \item Approx 3:
evaluation of the horizon distance neglecting the redshift
effects included in Eqns.~(\ref{SNRdef}) and (\ref{redf}).
\end{enumerate} 
In what follows, we
investigate the effects of  these approximations;
since in \cite{AbaAbbAbb2010b} they are used all together, we will show 
the consequences of applying approximation 1,  1+2  and 1+2+3 .
\begin{figure*}
 \centering
 \includegraphics[width=8.5cm,height=6cm]{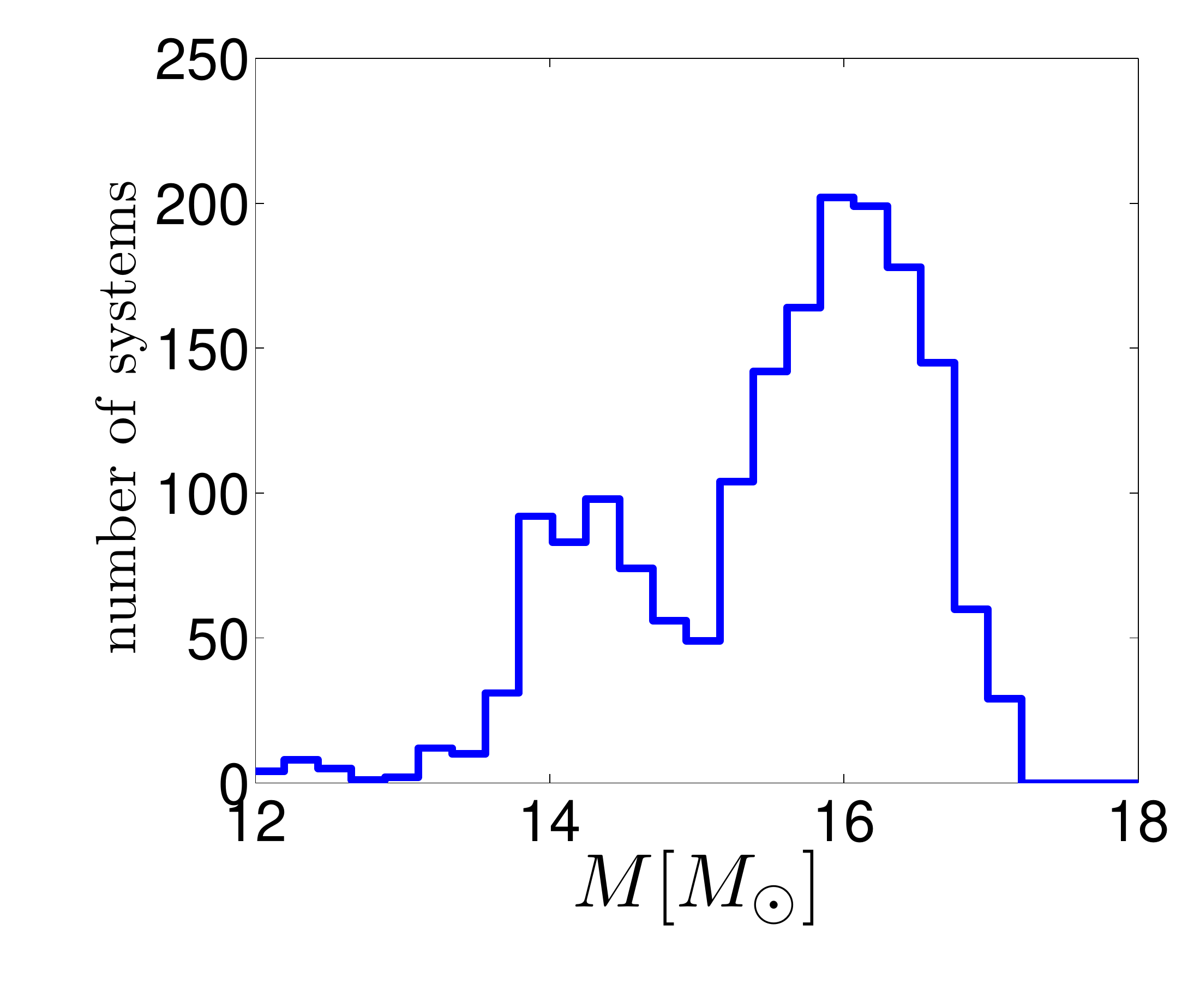}
 \includegraphics[width=8.5cm,height=6cm]{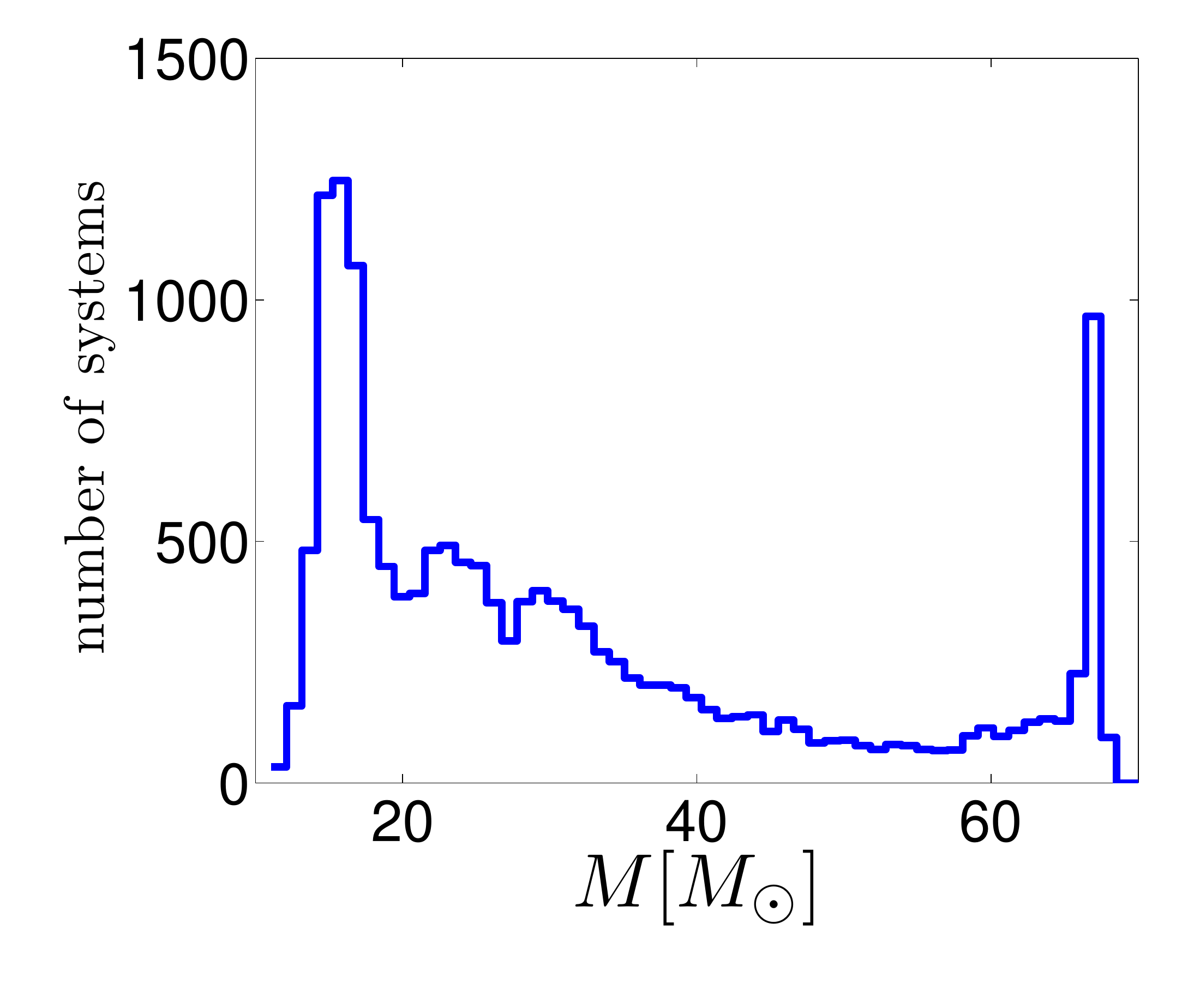}\\
 \includegraphics[width=8.5cm,height=6cm]{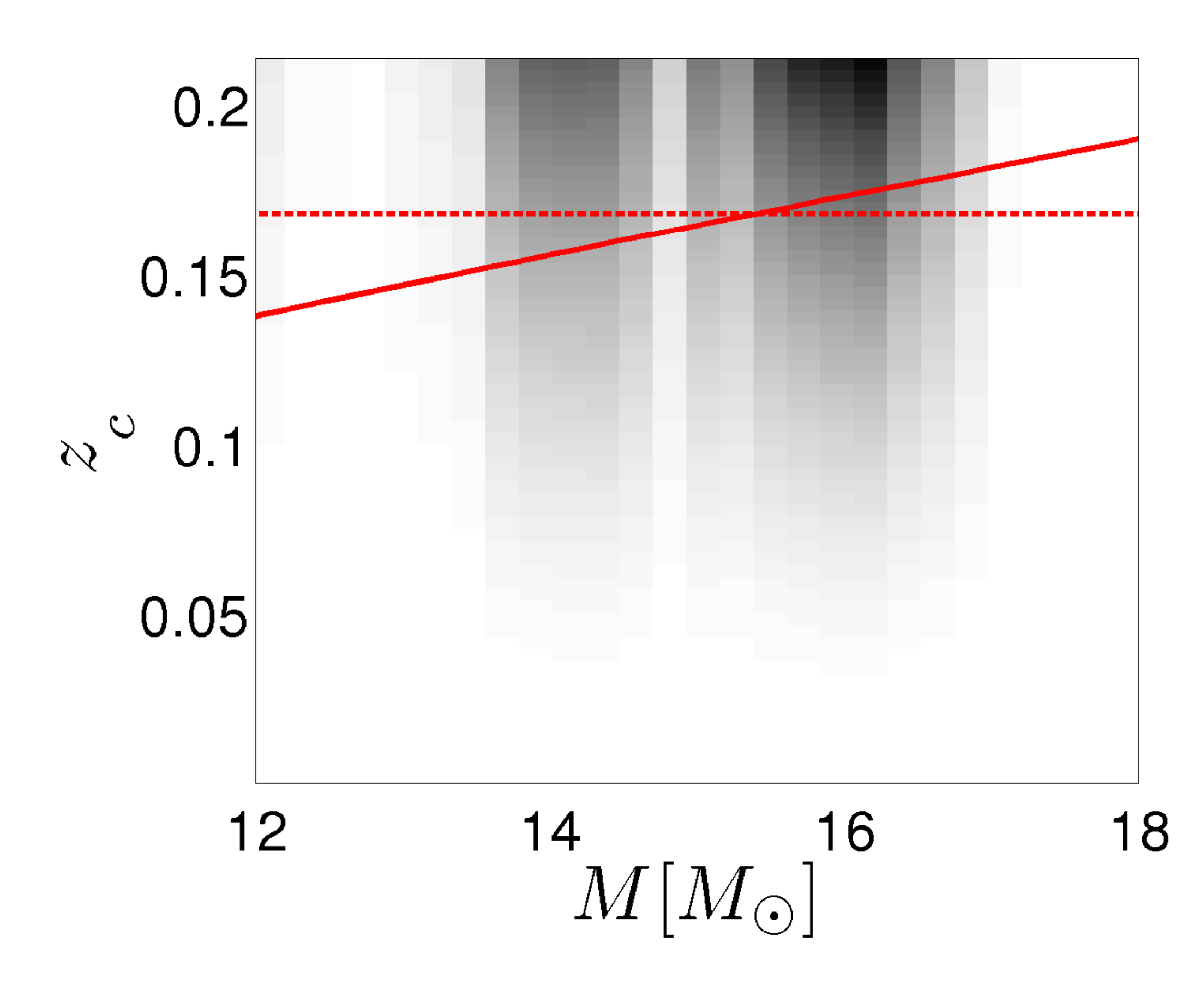}
 \includegraphics[width=8.5cm,height=6cm]{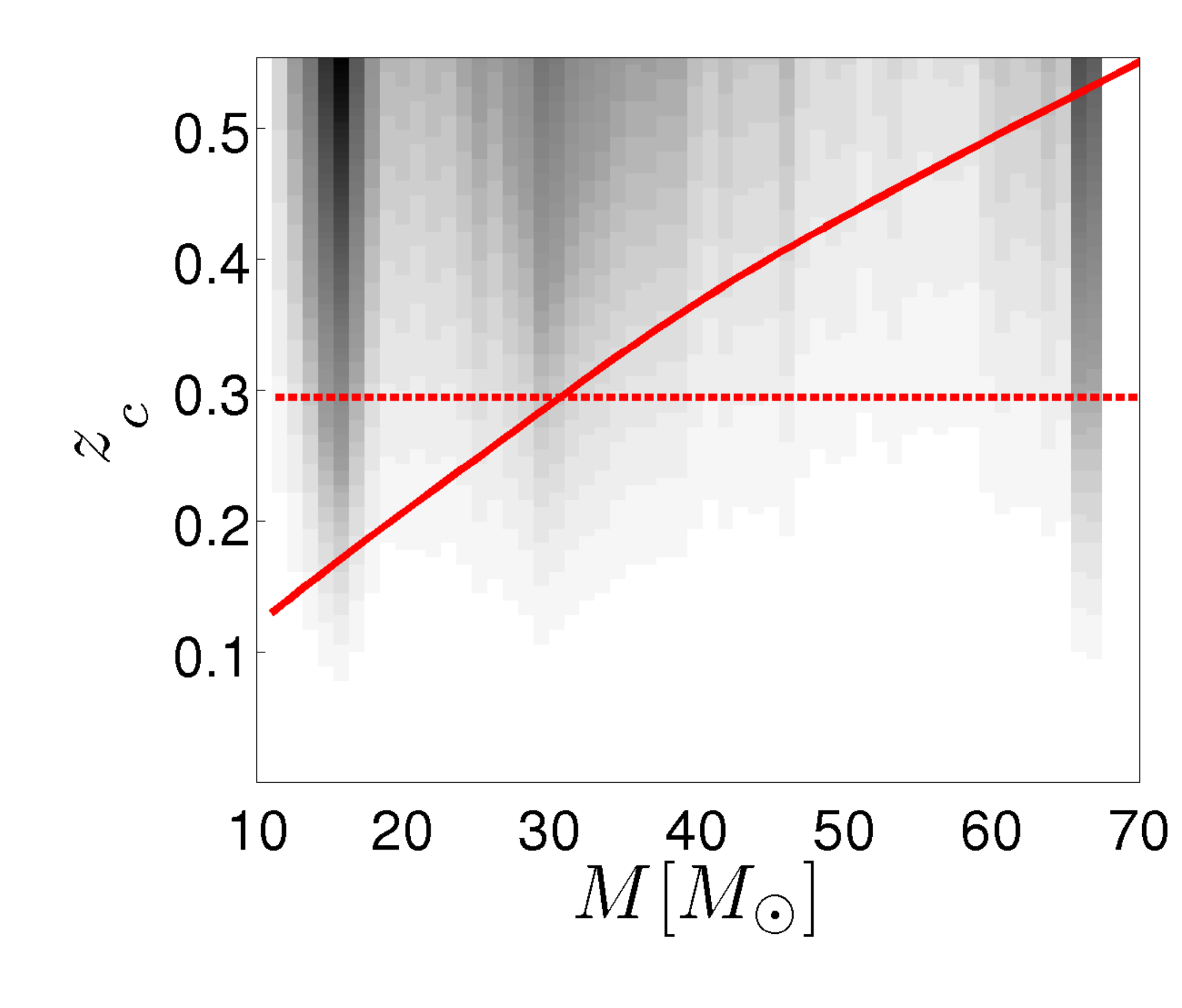}
\caption{
The total mass distribution for BH-BH systems for model A (left) and B
(right) is plotted in the upper panels.
In the lower panels we plot the integrand of Eq.~(\ref{Napprox1})
as a function of $M$ and redshift $z_{c}$, using a gray scale
proportional to its value.
The dashed line corresponds to $z_{\rm{max}}(\bar{M},\bar{\eta})$,  the solid
line corresponds to the redshift up to which the system with a given
total mass and $\eta=\bar{\eta}$ would be detected with an average SNR of 8.
}
\label{FIG2}
\end{figure*}

\subsection{Approximation 1} \label{sec:approx1} 
Approximation 1 consists in using the same horizon distance for all
systems of a binary population,
regardless of their individual total mass and symmetric mass ratio. To 
evaluate the effect of this approximation, we use the values of
$\bar{D}_{\rm h}$ given in Table~\ref{table2}, which 
correspond to the average values of $M$ and $\eta$
given in Eqns.~(\ref{mod1}) and
(\ref{mod2}).
We then compute the detection rate $N_{\rm{det}}^1$, by 
integrating Eq.~(\ref{Napprox0}) up to the redshift $z_{\rm max}$
\begin{equation} 
N_{\rm{det}}^{1} =
 \int dM
\int_{0}^{z_{\rm{max}}}
dz_c
\int 
\dot{r}_{\rm{coal}}
(M,\eta,z_c)\frac{dV}{dz_c}~d\eta~,
\label{Napprox1}
\end{equation}
\noindent
where $z_{\rm{max}}= z_{\rm{max}}(\bar{M},\bar{\eta})$.
The results are shown in Table~\ref{table3}. For model A,  
the difference between $N_{\rm{det}}^{1}$ and the no-approx rates
$N_{\rm{det}}$ is of a few percent, for all binary
populations, and  $N_{\rm{det}}^{1}$ is always larger than 
$N_{\rm{det}}$.
 
For model B, conversely, the difference between
$N_{\rm{det}}^{1}$ and $N_{\rm{det}}$  ranges between -26\% (ALIGO4) and
-36\% (ALIGO3) for BH-BH,  and between +13\% (ALIGO4) and
+19\% (AVirgo) for BH-NS. For NS-NS binaries the difference between
$N_{\rm{det}}^{1}$ and $N_{\rm{det}}$  is, as for  model A,
of a few percent for all interferometers. Moreover, 
for BH-BH and NS-NS families, $N_{\rm{det}}^{1}$ is systematically smaller
than $N_{\rm{det}}$, while it is always larger for BH-NS.

These data show that the effect of Approx 1  depends on the
model and on the kind of binary population. To clarify this dependence,
which will affect also the results obtained with Approx 2 and 3, 
in the upper panel of  Fig.~\ref{FIG2} we plot, as an example,  the
total mass distribution for BH-BH systems for model A (left) and B
(right).
In the lower panels  we plot the function
\be
\int 
\dot{r}_{\rm{coal}} (M,\eta,z_c)\frac{dV}{dz_c}~d\eta~,
\label{integrand}
\ee
which appear in Eq.~(\ref{Napprox1}),
as a function of $M$ and of the  redshift $z_{c}$,
using a gray scale proportional to its value. 
In these panels, the
horizontal dashed line corresponds to $z_{\rm{max}}(\bar{M},\bar{\eta})$, whereas the solid
line represent the function $z_{\rm{max}}(M,\bar{\eta})$, i.e. the redshift up to 
which the system with total mass $M$ and symmetric mass ratio
$\bar{\eta}$
would be detected with an average SNR of 8.
When the function (\ref{integrand}) is integrated up to $\bar{z}_{\rm{h}}$ to
compute the rate as in Eq.~(\ref{Napprox1}),  the
contribution of masses lower than the crossing point between the two
lines is overestimated, whereas that of masses
above that point is underestimated, with respect to their contribution
in Eq.~(\ref{Napprox0}). 
\begin{table*}[ht]
  \begin{center}
    \begin{tabular}{|c|c|c|c|c|c|c|c|c|c|c|c|c|}
      \hline
      \multicolumn{13}{|c|}{\multirow{2}*{Approx 1+2: detection
rates $[\rm{yr}^{-1}]$}}\\
      \multicolumn{13}{|c|}{} \\
      \hline
& \multicolumn{6}{|c|}{model A}&\multicolumn{6}{|c|}{model B}\\
      \cline{2-13}
      \multirow{2}*{detector}   & \multicolumn{2}{|c|}{BH-BH}
& \multicolumn{2}{|c|}{NS-NS}           & \multicolumn{2}{|c|}{BH-NS}
&\multicolumn{2}{|c|}{BH-BH}
& \multicolumn{2}{|c|}{NS-NS}           & \multicolumn{2}{|c|}{BH-NS}
 \\
      \cline{2-13}
                                & \multirow{2}*{$N_{\rm{det}}^{12}$} &
\multirow{2}*{$N_{\rm{det}}$} & \multirow{2}*{$N_{\rm{det}}^{12}$} &
\multirow{2}*{$N_{\rm{det}}$} &
                                  \multirow{2}*{$N_{\rm{det}}^{12}$} &
\multirow{2}*{$N_{\rm{det}}$} 
& \multirow{2}*{$N_{\rm{det}}^{12}$} &
\multirow{2}*{$N_{\rm{det}}$} & \multirow{2}*{$N_{\rm{det}}^{12}$} &
\multirow{2}*{$N_{\rm{det}}$} &
                                  \multirow{2}*{$N_{\rm{det}}^{12}$} &
\multirow{2}*{$N_{\rm{det}}$} 
\\
      & & & & & &  & & & & & &\\
      \hline
	AVirgo  & 91   & 73   & 2.7  & 2.3  & 4.6  & 3.7
&$5.62\cdot10^3$& $5.67\cdot10^3$ & 0.94&0.82 & 9.20 & 6.61\\	
      \hline
	ALIGO1 	& 273  & 203  & 6.9  & 6.0  & 12.1 & 9.2
&$1.73\cdot10^4$& $1.52\cdot10^4$ & 2.46 & 2.10 & 24.7 & 17.8\\	
      \hline
	ALIGO2	& 173  & 134  & 4.4  & 3.9  & 7.9  & 6.2
&$1.09\cdot10^4$& $1.07\cdot10^4$ & 1.56 & 1.35 & 16.4 &12.0\\	
      \hline
	ALIGO3  & 134  & 105  & 3.5  & 3.1  & 6.3  
&5.0&$8.09\cdot10^3$&$8.34\cdot10^3$ & 1.25 & 1.08& 13.0 & 9.60 \\	
      \hline
	ALIGO4 	& 347  & 254  & 8.8  & 7.6  & 15.7 
&11.8&$2.30\cdot10^4$& $1.84\cdot10^4$ &3.12 & 2.66 & 31.5 & 22.0\\	
      \hline
    \end{tabular}
\caption{ The detection rate $N_{\rm{det}}^{12}$, evaluated from
Eq.~(\ref{det2}) using Approx 1+2, for model A and B.
$N_{\rm{det}}^{12}$ for BH-BH systems (column 2), is
compared to the no-approx rates, $N_{\rm{det}}$, computed from
Eq.~(\ref{Napprox0}) (column 3).  The additional columns refer to the
same quantities evaluated for NS-NS and BH-NS  binaries.}
\label{table4} 
\end{center} 
\end{table*}

\begin{table*}[ht]
  \begin{center}
    \begin{tabular}{|c|c|c|c|c|c|c|c|c|c|c|c|c|}
      \hline
      \multicolumn{13}{|c|}{\multirow{2}*{Horizon distances [Mpc]}}\\
      \multicolumn{13}{|c|}{} \\
      \hline
		& \multicolumn{6}{|c|}{model A} & \multicolumn{6}{|c|}{model B} \\
      \cline{2-13}
      \multirow{2}*{detector}   & \multicolumn{2}{|c|}{BH-BH}  &
\multicolumn{2}{|c|}{NS-NS} & \multicolumn{2}{|c|}{BH-NS}
                                & \multicolumn{2}{|c|}{BH-BH}  &
\multicolumn{2}{|c|}{NS-NS} &
\multicolumn{2}{|c|}{BH-NS}\\
      \cline{2-13}
		& \multirow{2}*{$\bar{D}^3_{\rm{h}}$} & \multirow{2}*{$\bar{D}_{\rm{h}}$}
		& \multirow{2}*{$\bar{D}^3_{\rm{h}}$} & \multirow{2}*{$\bar{D}_{\rm{h}}$}
		& \multirow{2}*{$\bar{D}^3_{\rm{h}}$} & \multirow{2}*{$\bar{D}_{\rm{h}}$}
		& \multirow{2}*{$\bar{D}^3_{\rm{h}}$} & \multirow{2}*{$\bar{D}_{\rm{h}}$}
		& \multirow{2}*{$\bar{D}^3_{\rm{h}}$} & \multirow{2}*{$\bar{D}_{\rm{h}}$}
		& \multirow{2}*{$\bar{D}^3_{\rm{h}}$} &
\multirow{2}*{$\bar{D}_{\rm{h}}$} \\
      & & & & & & & & & & & & \\
      \hline
      AVirgo   	& 1540 & 1980 & 347 & 371 & 830  & 959  & 2650 & 4030 &
356 & 381 & 829  & 955  \\
      ALIGO1  	& 2120 & 2980 & 474 & 518 & 1140 & 1390 & 3650 & 6110 & 486
& 533 & 1140 & 1390 \\
      ALIGO2  	& 1860 & 2520 & 412 & 445 & 1000 & 1200 & 3200 & 5210 & 422
& 458 & 1000 & 1200 \\
      ALIGO3   	& 1720 & 2290 & 383 & 412 & 931  & 1100 & 2960 & 4580 &
392 & 423 & 934  & 1100 \\
      ALIGO4   	& 2280 & 3260 & 511 & 562 & 1230 & 1510 & 3920 & 6770 &
524 & 578 & 1230 & 1500 \\
      \hline
    \end{tabular}
    \caption{Average horizon distances (in Mpc) for the three families of
binaries
and for models A and B. $\bar{D}_{\rm{h}}^3$ is computed neglecting
redshift contributions, whereas these are included in
$\bar{D}_{\rm{h}}$. Both quantities are evaluated using the mean
values $\bar{M}$ and $\bar{\eta}$ given in Eqns.~(\ref{mod1})
and~(\ref{mod2}).}
    \label{table5}
  \end{center}
\end{table*}
From Figure~\ref{FIG2}, we see that for model A, the two contributions
nearly balance, whereas  
for model B, which has  a bimodal  mass distribution,
the high mass contribution, which  is relevant, is
neglected.
This explains why, for the BH-BH population of model B, 
$N_{\rm{det}}^{1}$ is systematically
lower than $N_{\rm{det}}$. A similar explanation holds for the BH-NS
population of model B, for which  the low mass contributions is larger
than the high mass one.
Thus, the rate estimated using Approx 1 can either be larger or
smaller than $N_{\rm{det}}$ by a different amount, depending on shape of
the mass
distribution. It has to be noted that the effect of this approximation
is not related
to the detector sensitivity because, for example, for model B the
largest
effect is for ALIGO3, which does not have the largest horizon distance
(see
Table~\ref{table2}).
\subsection{Approximation 1+2} \label{sec:approx2}
Approximation 2 assumes that the
coalescence rate $\dot{\rho}_c$ is constant through the whole detection
volume $V_{\rm{h}}$, and equal to the value at $z=0$ (Eqns.~(\ref{det1})
or (\ref{det2})).  As a consequence, the dependence of the
coalescence rate on redshift is completely neglected.

In addition, the detection rate is obtained by multiplying 
$\dot{\rho}^{\rm{local}}_c=\dot{\rho}_c(z=0)$ by the 
Euclidean volume $V_{\rm h}$ (Eq.~(\ref{Vhor}))
within $D_{\rm max}(\bar{M},\bar{\eta})$.

To assess the bias introduced  using  these approximations, we compute the
detection rates, $N_{\rm{det}}^{12}$, as:
\begin{equation}
 N_{\rm{det}}^{12} = \frac{4}{3}\pi\ D_{\rm max}(\bar{M},\bar{\eta})^3 \times
					\rho_c^{\rm local} \label{Napprox12}
\end{equation}
The results are given
in Table~\ref{table4}, where these are compared to the detection rates
$N_{\rm{det}}$ computed in
section~\ref{sec:ratecomp} with no approximation.

Let us consider model A first. 
The values of $N_{\rm{det}}^{12}$ are always larger than those of
$N_{\rm{det}}$, i.e. Approx 1+2 overestimates the rates.
The most significant differences are for BH-BH, and 
vary from +24\% (AVirgo)  to +37\% (ALIGO4); 
for NS-NS they vary  from +13\% (ALIGO2,ALIGO3) to +17\% (AVirgo) and for
BH-NS from +24\% (AVirgo) to +33\% (ALIGO4).
Since, as discussed above, for model A Approx 1
does not introduce a significant bias in the estimated rates,
it is clear that the differences are due mainly to Approx 2
alone.

These can be traced back to the contribution of the two assumptions:
the first is  that $\dot{\rho}_c$ is assumed to be constant
and equal to the $z=0$ value, while it is clear from
Fig.~\ref{FIG1}, that this is not the case.  
This assumption alone would lead to an underestimate of the
rates for all families because, in the range of interest for $z$,
the function  $\dot{\rho}_c(z)$ is always larger than the local value.  
\begin{table*}[ht]
  \begin{center}
\begin{tabular}{|c|c|c|c|c|c|c|c|c|c|c|c|c|}
      \hline
      \multicolumn{13}{|c|}{\multirow{2}*{Approx 1+2+3: detection
rates $[\rm{yr}^{-1}]$}}\\
      \multicolumn{13}{|c|}{} \\
      \hline
& \multicolumn{6}{|c|}{model A}&\multicolumn{6}{|c|}{model B}\\
      \cline{2-13}
      \multirow{2}*{detector}   & \multicolumn{2}{|c|}{BH-BH}
& \multicolumn{2}{|c|}{NS-NS}           & \multicolumn{2}{|c|}{BH-NS}
&\multicolumn{2}{|c|}{BH-BH}
& \multicolumn{2}{|c|}{NS-NS}           & \multicolumn{2}{|c|}{BH-NS}
 \\
      \cline{2-13}
                                & \multirow{2}*{$N_{\rm{det}}^{123}$} &
\multirow{2}*{$N_{\rm{det}}$} & \multirow{2}*{$N_{\rm{det}}^{123}$} &
\multirow{2}*{$N_{\rm{det}}$} &
                                  \multirow{2}*{$N_{\rm{det}}^{123}$} &
\multirow{2}*{$N_{\rm{det}}$}
& \multirow{2}*{$N_{\rm{det}}^{123}$} &
\multirow{2}*{$N_{\rm{det}}$} & \multirow{2}*{$N_{\rm{det}}^{123}$} &
\multirow{2}*{$N_{\rm{det}}$} &
                                  \multirow{2}*{$N_{\rm{det}}^{123}$} &
\multirow{2}*{$N_{\rm{det}}$}
\\
      & & & & & &  & & & & & &\\
      \hline
	AVirgo 	& 63.3 & 73   & 2.4  & 2.3  & 3.7  & 3.7
&$3.0\cdot10^3$& $5.67\cdot10^3$ & 0.85 &0.82   & 7.43 & 6.61\\	
      \hline
	ALIGO1 	& 166  & 203  & 6.1  & 6.0  & 9.0  & 9.2 
&$7.85\cdot10^3$& $1.52\cdot10^4$ & 2.15  & 2.10 & 18.4 & 17.8\\	
      \hline
	ALIGO2 	& 111  & 134  & 3.9  & 3.9  & 6.1  & 6.2  
&$5.31\cdot10^3$& $1.07\cdot10^4$ & 1.40 & 1.35  & 12.5 &12.0\\	
      \hline
	ALIGO3 	& 88.9 & 105  & 3.2  & 3.1  & 4.9  & 5.0  
&$4.21\cdot10^3$&$8.34\cdot10^3$ & 1.12 & 1.08   & 10.1 & 9.60 \\	
      \hline
	ALIGO4 	& 204  & 254  & 7.7  & 7.6  & 11.5 & 11.8 
&$9.76\cdot10^3$& $1.84\cdot10^4$ & 2.71 & 2.66  & 23.2 & 22.0\\	
      \hline
    \end{tabular}
    \caption{The detection rates 
 computed using Approx 1+2+3, i.e.
neglecting the effect of redshift, and assuming that the merger
rate is constant through the detection volume, and equal to its local
value.
The detection rates $N_{\rm{det}}$, computed with the no-approx
procedure (Eq.~(\ref{Napprox0})) are given for comparison.}
\label{table6}

  \end{center}
\end{table*}
Conversely, the  second assumption, i.e. that
the local coalescence  rate is multiplied by the Euclidean volume,
leads to an overestimate of the rates, because the Euclidean volume
is larger than the  cosmological one, and this effect  is
larger than the former, so that the combined result is an overestimate
of the detection rates for all the three families.

For model B, the detection rates $N_{\rm{det}}^{12}$ given in
Table~\ref{table4}, differ (in percentual) from the 
corresponding values of $N_{\rm{det}}$ as follows:
for BH-BH from -3\% (ALIGO3) to +25\% (ALIGO4), for NS-NS from +15\%
(AVirgo) to +17\%
(ALIGO1,ALIGO4), for BH-NS from +35\% (ALIGO3) to +43\% (ALIGO4). By comparing
$N_{\rm det}^1$ in table~\ref{table3} with $N_{\rm det}^{12}$ in table~\ref{table4}
we see that also in this case the effect of Approx 2 alone is to raise the
rates obtained with Approx 1 alone.
It should be noted that, contrary to the results of Approx 1, the effect
of Approx 2 depends on the reach of the interferometer, and is larger for 
interferometers with higher sensitivity. This is particularly evident for 
model A, where the effects of Approx 1 are negligible, while those
of Approx 1+2 increase going from AVirgo to ALIGO4, which has the highest
sensitivity.

For model B the effect of Approx 1 is not negligible, so that the effect of
Approx 1+2 varies between interferometers and binary families.
\subsection{Approximation 1+2+3} \label{sec:approx3}

Approximation 3 does not include redshift contributions in the
SNR evaluation. This will reflect in different values for $D_{\rm h}$ and
$D_{\rm max}$ which we will denote as $D_{\rm h}^3$ and $D_{\rm max}^3$.
In Table~\ref{table5} we compare the horizon distances $\bar{D}_{\rm h}$
given in Table~\ref{table2}, and which include redshift contribution,
to $\bar{D}_{\rm h}^3\equiv D_{\rm h}^3(\bar{M},\bar{\eta})$.

Horizon distances computed neglecting the redshift are smaller than
those which include it, and this reflects into smaller detection rates,
as can be seen  by comparing the rates $N_{\rm{det}}^{12}$ in
table~\ref{table4} for approx 1+2, with $N_{\rm{det}}^{123}$, given  in
Table~\ref{table6} and computed for Approx 1+2+3.  As expected, the
difference is larger for interferometers with larger horizon distances.

Turning to the comparison between  $N_{\rm{det}}^{123}$ and
$N_{\rm{det}}$, shown in Table~\ref{table6}, we see that the percentage
differences for the
three families of binaries range within:
for model A, from -13\% (AVirgo) to -18\%  (ALIGO1) for BH-BH,
from 0 (ALIGO2) to +4\% (AVirgo) for NS-NS, and  from 0 (AVirgo) to -3\%
(ALIGO4) for BH-NS;
for model B, from -47\% (AVirgo) to -51\% (ALIGO2) for BH-BH, from +2\%
(ALIGO1,ALIGO4) to +4\% (AVirgo,ALIGO2,ALIGO3) for NS-NS, from +3\%
(ALIGO1) to +12\% (AVirgo) for BH-NS.

Thus, it is clear that  Approx 1+2+3, which has been used to estimate the
detection rates of second generation detectors in
\cite{BelBulBai2011,BelDomBul2010,BelTaaKal2007, OshKimKal2008},
can result in an overestimate or underestimate of the rates depending 
on the details of the population and  on the interferometer which is  considered.

\section{Concluding remarks}
\label{sec:conclusions}
In this paper we discuss the procedure to compute the
detection rates for coalescing compact binaries, to be detected by
the  advanced  detectors expected to be in operation in a few years.  We
quantify the effect of various approximations adopted in the literature
to estimate these rates, and show that they introduce a
bias which depends not only on the approximation which is used, but also
on the statistical properties of the considered binary population
(BH-BH, NS-NS, or BH-NS).  These properties depend on the physical
inputs introduced in the population synthesis codes used to generate
the population, and these may vary significantly from one model to another.  
An example is shown in Figure~\ref{FIG2}, where the difference between the mass
distribution of BH-BH binaries for model A, which assumes solar
metallicity, and model B, based on subsolar metallicity, is apparent.

Thus, if we want to  associate a detection rate to 
a given population of binaries, derived on the basis of some physical
assumptions,  we need to use the fully consistent procedure  described in
section~\ref{sec:ratecomp}; 
indeed, the use of approximations as those described in 
section~\ref{sec:comparison},  would not
allow to disentangle the effects of the physical inputs from those of the assumed
approximations. Finally, we give an estimate of the detection rates for
the Einstein Telescope, ET, using 
two population models obtained by the updated version of the
{\tt Startrack} code, which are publicly available. 
These rates have been obtained using  the no approximation  procedure, 
which is the only one applicable to ET, due to its high sensitivity.

\section*{Acknowledgments}
We would like to thank the {\tt Startrack} collaboration, 
for making the results of {\tt Startrack} available to the community
on website {\tt http://www.syntheticuniverse.org}. 
\label{lastpage}

\bibliographystyle{apsrev}

\begin{thebibliography}{32}
\expandafter\ifx\csname natexlab\endcsname\relax\def\natexlab#1{#1}\fi
\expandafter\ifx\csname bibnamefont\endcsname\relax
  \def\bibnamefont#1{#1}\fi
\expandafter\ifx\csname bibfnamefont\endcsname\relax
  \def\bibfnamefont#1{#1}\fi
\expandafter\ifx\csname citenamefont\endcsname\relax
  \def\citenamefont#1{#1}\fi
\expandafter\ifx\csname url\endcsname\relax
  \def\url#1{\texttt{#1}}\fi
\expandafter\ifx\csname urlprefix\endcsname\relax\def\urlprefix{URL }\fi
\providecommand{\bibinfo}[2]{#2}
\providecommand{\eprint}[2][]{\url{#2}}

\bibitem[{\citenamefont{{Punturo} et~al.}(2010)\citenamefont{{Punturo},
  {Abernathy}, {Acernese}, {Allen}, {Andersson}, {Arun}, {Barone}, {Barr},
  {Barsuglia}, {Beker} et~al.}}]{PunAbeAce2010}
\bibinfo{author}{\bibfnamefont{M.}~\bibnamefont{{Punturo}}},
  \bibinfo{author}{\bibfnamefont{M.}~\bibnamefont{{Abernathy}}},
  \bibinfo{author}{\bibfnamefont{F.}~\bibnamefont{{Acernese}}},
  \bibinfo{author}{\bibfnamefont{B.}~\bibnamefont{{Allen}}},
  \bibinfo{author}{\bibfnamefont{N.}~\bibnamefont{{Andersson}}},
  \bibinfo{author}{\bibfnamefont{K.}~\bibnamefont{{Arun}}},
  \bibinfo{author}{\bibfnamefont{F.}~\bibnamefont{{Barone}}},
  \bibinfo{author}{\bibfnamefont{B.}~\bibnamefont{{Barr}}},
  \bibinfo{author}{\bibfnamefont{M.}~\bibnamefont{{Barsuglia}}},
  \bibinfo{author}{\bibfnamefont{M.}~\bibnamefont{{Beker}}},
  \bibnamefont{et~al.}, \bibinfo{journal}{Classical and Quantum Gravity}
  \textbf{\bibinfo{volume}{27}}, \bibinfo{pages}{194002}
  (\bibinfo{year}{2010}).

\bibitem[{\citenamefont{{Regimbau} et~al.}(2012)\citenamefont{{Regimbau},
  {Dent}, {Del Pozzo}, {Giampanis}, {Li}, {Robinson}, {Van Den Broeck},
  {Meacher}, {Rodriguez}, {Sathyaprakash} et~al.}}]{RegDenDel2012}
\bibinfo{author}{\bibfnamefont{T.}~\bibnamefont{{Regimbau}}},
  \bibinfo{author}{\bibfnamefont{T.}~\bibnamefont{{Dent}}},
  \bibinfo{author}{\bibfnamefont{W.}~\bibnamefont{{Del Pozzo}}},
  \bibinfo{author}{\bibfnamefont{S.}~\bibnamefont{{Giampanis}}},
  \bibinfo{author}{\bibfnamefont{T.~G.~F.} \bibnamefont{{Li}}},
  \bibinfo{author}{\bibfnamefont{C.}~\bibnamefont{{Robinson}}},
  \bibinfo{author}{\bibfnamefont{C.}~\bibnamefont{{Van Den Broeck}}},
  \bibinfo{author}{\bibfnamefont{D.}~\bibnamefont{{Meacher}}},
  \bibinfo{author}{\bibfnamefont{C.}~\bibnamefont{{Rodriguez}}},
  \bibinfo{author}{\bibfnamefont{B.~S.} \bibnamefont{{Sathyaprakash}}},
  \bibnamefont{et~al.}, \bibinfo{journal}{ArXiv e-prints}
  (\bibinfo{year}{2012}), \eprint{1201.3563}.

\bibitem[{\citenamefont{{Sathyaprakash}
  et~al.}(2012)\citenamefont{{Sathyaprakash}, {Abernathy}, {Acernese}, {Ajith},
  {Allen}, {Amaro-Seoane}, {Andersson}, {Aoudia}, {Arun}, {Astone}
  et~al.}}]{SatAbeAce2012}
\bibinfo{author}{\bibfnamefont{B.}~\bibnamefont{{Sathyaprakash}}},
  \bibinfo{author}{\bibfnamefont{M.}~\bibnamefont{{Abernathy}}},
  \bibinfo{author}{\bibfnamefont{F.}~\bibnamefont{{Acernese}}},
  \bibinfo{author}{\bibfnamefont{P.}~\bibnamefont{{Ajith}}},
  \bibinfo{author}{\bibfnamefont{B.}~\bibnamefont{{Allen}}},
  \bibinfo{author}{\bibfnamefont{P.}~\bibnamefont{{Amaro-Seoane}}},
  \bibinfo{author}{\bibfnamefont{N.}~\bibnamefont{{Andersson}}},
  \bibinfo{author}{\bibfnamefont{S.}~\bibnamefont{{Aoudia}}},
  \bibinfo{author}{\bibfnamefont{K.}~\bibnamefont{{Arun}}},
  \bibinfo{author}{\bibfnamefont{P.}~\bibnamefont{{Astone}}},
  \bibnamefont{et~al.}, \bibinfo{journal}{eprint arXiv:1206.0331}
  (\bibinfo{year}{2012}), \eprint{1206.0331}.

\bibitem[{\citenamefont{{Sathyaprakash}
  et~al.}(2010)\citenamefont{{Sathyaprakash}, {Schutz}, and {Van Den
  Broeck}}}]{SatSchVan2010}
\bibinfo{author}{\bibfnamefont{B.~S.} \bibnamefont{{Sathyaprakash}}},
  \bibinfo{author}{\bibfnamefont{B.~F.} \bibnamefont{{Schutz}}},
  \bibnamefont{and} \bibinfo{author}{\bibfnamefont{C.}~\bibnamefont{{Van Den
  Broeck}}}, \bibinfo{journal}{Classical and Quantum Gravity, Volume 27, Issue
  21, pp.~215006 (2010).} \textbf{\bibinfo{volume}{27}},
  \bibinfo{pages}{215006} (\bibinfo{year}{2010}), \eprint{0906.4151}.

\bibitem[{\citenamefont{{Taylor} and {Gair}}(2012)}]{TayGai2012}
\bibinfo{author}{\bibfnamefont{S.~R.} \bibnamefont{{Taylor}}} \bibnamefont{and}
  \bibinfo{author}{\bibfnamefont{J.~R.} \bibnamefont{{Gair}}},
  \bibinfo{journal}{eprint arXiv:1204.6739}  (\bibinfo{year}{2012}),
  \eprint{1204.6739}.

\bibitem[{\citenamefont{{Taylor} et~al.}(2012)\citenamefont{{Taylor}, {Gair},
  and {Mandel}}}]{TayGaiMan2012}
\bibinfo{author}{\bibfnamefont{S.~R.} \bibnamefont{{Taylor}}},
  \bibinfo{author}{\bibfnamefont{J.~R.} \bibnamefont{{Gair}}},
  \bibnamefont{and} \bibinfo{author}{\bibfnamefont{I.}~\bibnamefont{{Mandel}}},
  \bibinfo{journal}{\prd} \textbf{\bibinfo{volume}{85}}, \bibinfo{eid}{023535}
  (\bibinfo{year}{2012}), \eprint{1108.5161}.

\bibitem[{\citenamefont{{Abadie} et~al.}(2011)\citenamefont{{Abadie}, {Abbott},
  {Abbott}, {Abernathy}, {Accadia}, {Acernese}, {Adams}, {Adhikari}, {Ajith},
  {Allen} et~al.}}]{AbaAbbAbb2011}
\bibinfo{author}{\bibfnamefont{J.}~\bibnamefont{{Abadie}}},
  \bibinfo{author}{\bibfnamefont{B.~P.} \bibnamefont{{Abbott}}},
  \bibinfo{author}{\bibfnamefont{R.}~\bibnamefont{{Abbott}}},
  \bibinfo{author}{\bibfnamefont{M.}~\bibnamefont{{Abernathy}}},
  \bibinfo{author}{\bibfnamefont{T.}~\bibnamefont{{Accadia}}},
  \bibinfo{author}{\bibfnamefont{F.}~\bibnamefont{{Acernese}}},
  \bibinfo{author}{\bibfnamefont{C.}~\bibnamefont{{Adams}}},
  \bibinfo{author}{\bibfnamefont{R.}~\bibnamefont{{Adhikari}}},
  \bibinfo{author}{\bibfnamefont{P.}~\bibnamefont{{Ajith}}},
  \bibinfo{author}{\bibfnamefont{B.}~\bibnamefont{{Allen}}},
  \bibnamefont{et~al.}, \bibinfo{journal}{\prd} \textbf{\bibinfo{volume}{83}},
  \bibinfo{eid}{122005} (\bibinfo{year}{2011}), \eprint{1102.3781}.

\bibitem[{\citenamefont{{Abadie} et~al.}(2012)\citenamefont{{Abadie}, {Abbott},
  {Abbott}, {Abbott}, {Abernathy}, {Accadia}, {Acernese}, {Adams}, {Adhikari},
  {Affeldt} et~al.}}]{AbaAbbAbb2012b}
\bibinfo{author}{\bibfnamefont{J.}~\bibnamefont{{Abadie}}},
  \bibinfo{author}{\bibfnamefont{B.~P.} \bibnamefont{{Abbott}}},
  \bibinfo{author}{\bibfnamefont{R.}~\bibnamefont{{Abbott}}},
  \bibinfo{author}{\bibfnamefont{T.~D.} \bibnamefont{{Abbott}}},
  \bibinfo{author}{\bibfnamefont{M.}~\bibnamefont{{Abernathy}}},
  \bibinfo{author}{\bibfnamefont{T.}~\bibnamefont{{Accadia}}},
  \bibinfo{author}{\bibfnamefont{F.}~\bibnamefont{{Acernese}}},
  \bibinfo{author}{\bibfnamefont{C.}~\bibnamefont{{Adams}}},
  \bibinfo{author}{\bibfnamefont{R.}~\bibnamefont{{Adhikari}}},
  \bibinfo{author}{\bibfnamefont{C.}~\bibnamefont{{Affeldt}}},
  \bibnamefont{et~al.}, \bibinfo{journal}{\prd} \textbf{\bibinfo{volume}{85}},
  \bibinfo{eid}{022001} (\bibinfo{year}{2012}), \eprint{1110.0208}.

\bibitem[{\citenamefont{{Abadie} et~al.}(2010)\citenamefont{{Abadie}, {Abbott},
  {Abbott}, {Abernathy}, {Accadia}, {Acernese}, {Adams}, {Adhikari}, {Ajith},
  {Allen} et~al.}}]{AbaAbbAbb2010b}
\bibinfo{author}{\bibfnamefont{J.}~\bibnamefont{{Abadie}}},
  \bibinfo{author}{\bibfnamefont{B.~P.} \bibnamefont{{Abbott}}},
  \bibinfo{author}{\bibfnamefont{R.}~\bibnamefont{{Abbott}}},
  \bibinfo{author}{\bibfnamefont{M.}~\bibnamefont{{Abernathy}}},
  \bibinfo{author}{\bibfnamefont{T.}~\bibnamefont{{Accadia}}},
  \bibinfo{author}{\bibfnamefont{F.}~\bibnamefont{{Acernese}}},
  \bibinfo{author}{\bibfnamefont{C.}~\bibnamefont{{Adams}}},
  \bibinfo{author}{\bibfnamefont{R.}~\bibnamefont{{Adhikari}}},
  \bibinfo{author}{\bibfnamefont{P.}~\bibnamefont{{Ajith}}},
  \bibinfo{author}{\bibfnamefont{B.}~\bibnamefont{{Allen}}},
  \bibnamefont{et~al.}, \bibinfo{journal}{Classical and Quantum Gravity}
  \textbf{\bibinfo{volume}{27}}, \bibinfo{pages}{173001}
  (\bibinfo{year}{2010}), \eprint{1003.2480}.

\bibitem[{\citenamefont{{Gair} et~al.}(2011)\citenamefont{{Gair}, {Mandel},
  {Miller}, and {Volonteri}}}]{GaiManMill2011}
\bibinfo{author}{\bibfnamefont{J.~R.} \bibnamefont{{Gair}}},
  \bibinfo{author}{\bibfnamefont{I.}~\bibnamefont{{Mandel}}},
  \bibinfo{author}{\bibfnamefont{M.~C.} \bibnamefont{{Miller}}},
  \bibnamefont{and}
  \bibinfo{author}{\bibfnamefont{M.}~\bibnamefont{{Volonteri}}},
  \bibinfo{journal}{General Relativity and Gravitation}
  \textbf{\bibinfo{volume}{43}}, \bibinfo{pages}{485} (\bibinfo{year}{2011}),
  \eprint{0907.5450}.

\bibitem[{\citenamefont{{Marassi} et~al.}(2011)\citenamefont{{Marassi},
  {Schneider}, {Corvino}, {Ferrari}, and {Zwart}}}]{MarSchCor2011}
\bibinfo{author}{\bibfnamefont{S.}~\bibnamefont{{Marassi}}},
  \bibinfo{author}{\bibfnamefont{R.}~\bibnamefont{{Schneider}}},
  \bibinfo{author}{\bibfnamefont{G.}~\bibnamefont{{Corvino}}},
  \bibinfo{author}{\bibfnamefont{V.}~\bibnamefont{{Ferrari}}},
  \bibnamefont{and} \bibinfo{author}{\bibfnamefont{S.~P.}
  \bibnamefont{{Zwart}}}, \bibinfo{journal}{\prd}
  \textbf{\bibinfo{volume}{84}}, \bibinfo{eid}{124037} (\bibinfo{year}{2011}),
  \eprint{1111.6125}.

\bibitem[{\citenamefont{{Mandel} and {O'Shaughnessy}}(2010)}]{ManOsh2010}
\bibinfo{author}{\bibfnamefont{I.}~\bibnamefont{{Mandel}}} \bibnamefont{and}
  \bibinfo{author}{\bibfnamefont{R.}~\bibnamefont{{O'Shaughnessy}}},
  \bibinfo{journal}{Classical and Quantum Gravity}
  \textbf{\bibinfo{volume}{27}}, \bibinfo{pages}{114007}
  (\bibinfo{year}{2010}), \eprint{0912.1074}.

\bibitem[{\citenamefont{{Bulik}
  et~al.}(2004{\natexlab{a}})\citenamefont{{Bulik}, {Belczy{\'n}ski}, and
  {Rudak}}}]{BulBelRud2004}
\bibinfo{author}{\bibfnamefont{T.}~\bibnamefont{{Bulik}}},
  \bibinfo{author}{\bibfnamefont{K.}~\bibnamefont{{Belczy{\'n}ski}}},
  \bibnamefont{and} \bibinfo{author}{\bibfnamefont{B.}~\bibnamefont{{Rudak}}},
  \bibinfo{journal}{A\&A} \textbf{\bibinfo{volume}{415}}, \bibinfo{pages}{407}
  (\bibinfo{year}{2004}{\natexlab{a}}), \eprint{arXiv:astro-ph/0307237}.

\bibitem[{\citenamefont{{Bulik}
  et~al.}(2004{\natexlab{b}})\citenamefont{{Bulik}, {Gondek-Rosinska}, and
  {Belczynski}}}]{BulGonBel2004}
\bibinfo{author}{\bibfnamefont{T.}~\bibnamefont{{Bulik}}},
  \bibinfo{author}{\bibfnamefont{D.}~\bibnamefont{{Gondek-Rosinska}}},
  \bibnamefont{and}
  \bibinfo{author}{\bibfnamefont{K.}~\bibnamefont{{Belczynski}}},
  \bibinfo{journal}{\mnras} \textbf{\bibinfo{volume}{352}},
  \bibinfo{pages}{1372} (\bibinfo{year}{2004}{\natexlab{b}}),
  \eprint{arXiv:astro-ph/0310544}.

\bibitem[{\citenamefont{{Voss} and {Tauris}}(2003)}]{VosTau2003}
\bibinfo{author}{\bibfnamefont{R.}~\bibnamefont{{Voss}}} \bibnamefont{and}
  \bibinfo{author}{\bibfnamefont{T.~M.} \bibnamefont{{Tauris}}},
  \bibinfo{journal}{\mnras} \textbf{\bibinfo{volume}{342}},
  \bibinfo{pages}{1169} (\bibinfo{year}{2003}),
  \eprint{arXiv:astro-ph/0303227}.

\bibitem[{\citenamefont{{O'Shaughnessy}
  et~al.}(2010)\citenamefont{{O'Shaughnessy}, {Kalogera}, and
  {Belczynski}}}]{OshKalBel2010}
\bibinfo{author}{\bibfnamefont{R.}~\bibnamefont{{O'Shaughnessy}}},
  \bibinfo{author}{\bibfnamefont{V.}~\bibnamefont{{Kalogera}}},
  \bibnamefont{and}
  \bibinfo{author}{\bibfnamefont{K.}~\bibnamefont{{Belczynski}}},
  \bibinfo{journal}{\apj} \textbf{\bibinfo{volume}{716}}, \bibinfo{pages}{615}
  (\bibinfo{year}{2010}), \eprint{0908.3635}.

\bibitem[{\citenamefont{{O'Shaughnessy}
  et~al.}(2008{\natexlab{a}})\citenamefont{{O'Shaughnessy}, {Belczynski}, and
  {Kalogera}}}]{OshBelKal2008}
\bibinfo{author}{\bibfnamefont{R.}~\bibnamefont{{O'Shaughnessy}}},
  \bibinfo{author}{\bibfnamefont{K.}~\bibnamefont{{Belczynski}}},
  \bibnamefont{and}
  \bibinfo{author}{\bibfnamefont{V.}~\bibnamefont{{Kalogera}}},
  \bibinfo{journal}{\apj} \textbf{\bibinfo{volume}{675}}, \bibinfo{pages}{566}
  (\bibinfo{year}{2008}{\natexlab{a}}), \eprint{0706.4139}.

\bibitem[{\citenamefont{{O'Shaughnessy}
  et~al.}(2008{\natexlab{b}})\citenamefont{{O'Shaughnessy}, {Kim}, {Kalogera},
  and {Belczynski}}}]{OshKimKal2008}
\bibinfo{author}{\bibfnamefont{R.}~\bibnamefont{{O'Shaughnessy}}},
  \bibinfo{author}{\bibfnamefont{C.}~\bibnamefont{{Kim}}},
  \bibinfo{author}{\bibfnamefont{V.}~\bibnamefont{{Kalogera}}},
  \bibnamefont{and}
  \bibinfo{author}{\bibfnamefont{K.}~\bibnamefont{{Belczynski}}},
  \bibinfo{journal}{\apj} \textbf{\bibinfo{volume}{672}}, \bibinfo{pages}{479}
  (\bibinfo{year}{2008}{\natexlab{b}}), \eprint{arXiv:astro-ph/0610076}.

\bibitem[{\citenamefont{{Belczynski} et~al.}(2011)\citenamefont{{Belczynski},
  {Bulik}, and {Bailyn}}}]{BelBulBai2011}
\bibinfo{author}{\bibfnamefont{K.}~\bibnamefont{{Belczynski}}},
  \bibinfo{author}{\bibfnamefont{T.}~\bibnamefont{{Bulik}}}, \bibnamefont{and}
  \bibinfo{author}{\bibfnamefont{C.}~\bibnamefont{{Bailyn}}},
  \bibinfo{journal}{\apjl} \textbf{\bibinfo{volume}{742}}, \bibinfo{eid}{L2}
  (\bibinfo{year}{2011}), \eprint{1107.4106}.

\bibitem[{\citenamefont{{Belczynski} et~al.}(2010)\citenamefont{{Belczynski},
  {Dominik}, {Bulik}, {O'Shaughnessy}, {Fryer}, and {Holz}}}]{BelDomBul2010}
\bibinfo{author}{\bibfnamefont{K.}~\bibnamefont{{Belczynski}}},
  \bibinfo{author}{\bibfnamefont{M.}~\bibnamefont{{Dominik}}},
  \bibinfo{author}{\bibfnamefont{T.}~\bibnamefont{{Bulik}}},
  \bibinfo{author}{\bibfnamefont{R.}~\bibnamefont{{O'Shaughnessy}}},
  \bibinfo{author}{\bibfnamefont{C.}~\bibnamefont{{Fryer}}}, \bibnamefont{and}
  \bibinfo{author}{\bibfnamefont{D.~E.} \bibnamefont{{Holz}}},
  \bibinfo{journal}{\apjl} \textbf{\bibinfo{volume}{715}},
  \bibinfo{pages}{L138} (\bibinfo{year}{2010}), \eprint{1004.0386}.

\bibitem[{\citenamefont{{Belczynski} et~al.}(2007)\citenamefont{{Belczynski},
  {Taam}, {Kalogera}, {Rasio}, and {Bulik}}}]{BelTaaKal2007}
\bibinfo{author}{\bibfnamefont{K.}~\bibnamefont{{Belczynski}}},
  \bibinfo{author}{\bibfnamefont{R.~E.} \bibnamefont{{Taam}}},
  \bibinfo{author}{\bibfnamefont{V.}~\bibnamefont{{Kalogera}}},
  \bibinfo{author}{\bibfnamefont{F.~A.} \bibnamefont{{Rasio}}},
  \bibnamefont{and} \bibinfo{author}{\bibfnamefont{T.}~\bibnamefont{{Bulik}}},
  \bibinfo{journal}{\apj} \textbf{\bibinfo{volume}{662}}, \bibinfo{pages}{504}
  (\bibinfo{year}{2007}), \eprint{arXiv:astro-ph/0612032}.

\bibitem[{\citenamefont{{O'Shaughnessy}
  et~al.}(2005)\citenamefont{{O'Shaughnessy}, {Kim}, {Fragos}, {Kalogera}, and
  {Belczynski}}}]{OshKimFra2005}
\bibinfo{author}{\bibfnamefont{R.}~\bibnamefont{{O'Shaughnessy}}},
  \bibinfo{author}{\bibfnamefont{C.}~\bibnamefont{{Kim}}},
  \bibinfo{author}{\bibfnamefont{T.}~\bibnamefont{{Fragos}}},
  \bibinfo{author}{\bibfnamefont{V.}~\bibnamefont{{Kalogera}}},
  \bibnamefont{and}
  \bibinfo{author}{\bibfnamefont{K.}~\bibnamefont{{Belczynski}}},
  \bibinfo{journal}{\apj} \textbf{\bibinfo{volume}{633}}, \bibinfo{pages}{1076}
  (\bibinfo{year}{2005}), \eprint{arXiv:astro-ph/0504479}.

\bibitem[{\citenamefont{{Dominik} et~al.}(2012)\citenamefont{{Dominik},
  {Belczynski}, {Fryer}, {Holz}, {Berti}, {Bulik}, {Mandel}, and
  {O'Shaughnessy}}}]{DomBelFry2012}
\bibinfo{author}{\bibfnamefont{M.}~\bibnamefont{{Dominik}}},
  \bibinfo{author}{\bibfnamefont{K.}~\bibnamefont{{Belczynski}}},
  \bibinfo{author}{\bibfnamefont{C.}~\bibnamefont{{Fryer}}},
  \bibinfo{author}{\bibfnamefont{D.}~\bibnamefont{{Holz}}},
  \bibinfo{author}{\bibfnamefont{E.}~\bibnamefont{{Berti}}},
  \bibinfo{author}{\bibfnamefont{T.}~\bibnamefont{{Bulik}}},
  \bibinfo{author}{\bibfnamefont{I.}~\bibnamefont{{Mandel}}}, \bibnamefont{and}
  \bibinfo{author}{\bibfnamefont{R.}~\bibnamefont{{O'Shaughnessy}}},
  \bibinfo{journal}{ArXiv e-prints}  (\bibinfo{year}{2012}),
  \eprint{1202.4901}.

\bibitem[{\citenamefont{{Flanagan} and {Hughes}}(1998)}]{FlaHug1998}
\bibinfo{author}{\bibfnamefont{{\'E}.~{\'E}.} \bibnamefont{{Flanagan}}}
  \bibnamefont{and} \bibinfo{author}{\bibfnamefont{S.~A.}
  \bibnamefont{{Hughes}}}, \bibinfo{journal}{\prd}
  \textbf{\bibinfo{volume}{57}}, \bibinfo{pages}{4535} (\bibinfo{year}{1998}),
  \eprint{arXiv:gr-qc/9701039}.

\bibitem[{\citenamefont{{Schneider} et~al.}(2001)\citenamefont{{Schneider},
  {Ferrari}, {Matarrese}, and {Portegies Zwart}}}]{SchFerMat2001}
\bibinfo{author}{\bibfnamefont{R.}~\bibnamefont{{Schneider}}},
  \bibinfo{author}{\bibfnamefont{V.}~\bibnamefont{{Ferrari}}},
  \bibinfo{author}{\bibfnamefont{S.}~\bibnamefont{{Matarrese}}},
  \bibnamefont{and} \bibinfo{author}{\bibfnamefont{S.~F.}
  \bibnamefont{{Portegies Zwart}}}, \bibinfo{journal}{\mnras}
  \textbf{\bibinfo{volume}{324}}, \bibinfo{pages}{797} (\bibinfo{year}{2001}),
  \eprint{arXiv:astro-ph/0002055}.

\bibitem[{\citenamefont{{Tornatore} et~al.}(2007)\citenamefont{{Tornatore},
  {Ferrara}, and {Schneider}}}]{TorFerSch2007}
\bibinfo{author}{\bibfnamefont{L.}~\bibnamefont{{Tornatore}}},
  \bibinfo{author}{\bibfnamefont{A.}~\bibnamefont{{Ferrara}}},
  \bibnamefont{and}
  \bibinfo{author}{\bibfnamefont{R.}~\bibnamefont{{Schneider}}},
  \bibinfo{journal}{\mnras} \textbf{\bibinfo{volume}{382}},
  \bibinfo{pages}{945} (\bibinfo{year}{2007}), \eprint{0707.1433}.

\bibitem[{\citenamefont{{Bouwens} et~al.}(2010)\citenamefont{{Bouwens},
  {Illingworth}, {Oesch}, {Stiavelli}, {van Dokkum}, {Trenti}, {Magee},
  {Labb{\'e}}, {Franx}, {Carollo} et~al.}}]{BouIllOes2010}
\bibinfo{author}{\bibfnamefont{R.~J.} \bibnamefont{{Bouwens}}},
  \bibinfo{author}{\bibfnamefont{G.~D.} \bibnamefont{{Illingworth}}},
  \bibinfo{author}{\bibfnamefont{P.~A.} \bibnamefont{{Oesch}}},
  \bibinfo{author}{\bibfnamefont{M.}~\bibnamefont{{Stiavelli}}},
  \bibinfo{author}{\bibfnamefont{P.}~\bibnamefont{{van Dokkum}}},
  \bibinfo{author}{\bibfnamefont{M.}~\bibnamefont{{Trenti}}},
  \bibinfo{author}{\bibfnamefont{D.}~\bibnamefont{{Magee}}},
  \bibinfo{author}{\bibfnamefont{I.}~\bibnamefont{{Labb{\'e}}}},
  \bibinfo{author}{\bibfnamefont{M.}~\bibnamefont{{Franx}}},
  \bibinfo{author}{\bibfnamefont{C.~M.} \bibnamefont{{Carollo}}},
  \bibnamefont{et~al.}, \bibinfo{journal}{\apjl}
  \textbf{\bibinfo{volume}{709}}, \bibinfo{pages}{L133} (\bibinfo{year}{2010}),
  \eprint{0909.1803}.

\bibitem[{\citenamefont{{Ajith} et~al.}(2011)\citenamefont{{Ajith}, {Hannam},
  {Husa}, {Chen}, {Br{\"u}gmann}, {Dorband}, {M{\"u}ller}, {Ohme}, {Pollney},
  {Reisswig} et~al.}}]{AjiHanHus2011}
\bibinfo{author}{\bibfnamefont{P.}~\bibnamefont{{Ajith}}},
  \bibinfo{author}{\bibfnamefont{M.}~\bibnamefont{{Hannam}}},
  \bibinfo{author}{\bibfnamefont{S.}~\bibnamefont{{Husa}}},
  \bibinfo{author}{\bibfnamefont{Y.}~\bibnamefont{{Chen}}},
  \bibinfo{author}{\bibfnamefont{B.}~\bibnamefont{{Br{\"u}gmann}}},
  \bibinfo{author}{\bibfnamefont{N.}~\bibnamefont{{Dorband}}},
  \bibinfo{author}{\bibfnamefont{D.}~\bibnamefont{{M{\"u}ller}}},
  \bibinfo{author}{\bibfnamefont{F.}~\bibnamefont{{Ohme}}},
  \bibinfo{author}{\bibfnamefont{D.}~\bibnamefont{{Pollney}}},
  \bibinfo{author}{\bibfnamefont{C.}~\bibnamefont{{Reisswig}}},
  \bibnamefont{et~al.}, \bibinfo{journal}{Physical Review Letters}
  \textbf{\bibinfo{volume}{106}}, \bibinfo{eid}{241101} (\bibinfo{year}{2011}),
  \eprint{0909.2867}.

\bibitem[{\citenamefont{{Harry} and {LIGO Scientific
  Collaboration}}(2010)}]{Harry2010}
\bibinfo{author}{\bibfnamefont{G.~M.} \bibnamefont{{Harry}}} \bibnamefont{and}
  \bibinfo{author}{\bibnamefont{{LIGO Scientific Collaboration}}},
  \bibinfo{journal}{Classical and Quantum Gravity}
  \textbf{\bibinfo{volume}{27}}, \bibinfo{pages}{084006}
  (\bibinfo{year}{2010}).

\bibitem[{\citenamefont{{Kopparapu} et~al.}(2008)\citenamefont{{Kopparapu},
  {Hanna}, {Kalogera}, {O'Shaughnessy}, {Gonz{\'a}lez}, {Brady}, and
  {Fairhurst}}}]{KopHanKal2008}
\bibinfo{author}{\bibfnamefont{R.~K.} \bibnamefont{{Kopparapu}}},
  \bibinfo{author}{\bibfnamefont{C.}~\bibnamefont{{Hanna}}},
  \bibinfo{author}{\bibfnamefont{V.}~\bibnamefont{{Kalogera}}},
  \bibinfo{author}{\bibfnamefont{R.}~\bibnamefont{{O'Shaughnessy}}},
  \bibinfo{author}{\bibfnamefont{G.}~\bibnamefont{{Gonz{\'a}lez}}},
  \bibinfo{author}{\bibfnamefont{P.~R.} \bibnamefont{{Brady}}},
  \bibnamefont{and}
  \bibinfo{author}{\bibfnamefont{S.}~\bibnamefont{{Fairhurst}}},
  \bibinfo{journal}{\apj} \textbf{\bibinfo{volume}{675}}, \bibinfo{pages}{1459}
  (\bibinfo{year}{2008}), \eprint{0706.1283}.

\bibitem[{\citenamefont{{Nelemans} et~al.}(2001)\citenamefont{{Nelemans},
  {Yungelson}, and {Portegies Zwart}}}]{NelYunPor2001a}
\bibinfo{author}{\bibfnamefont{G.}~\bibnamefont{{Nelemans}}},
  \bibinfo{author}{\bibfnamefont{L.~R.} \bibnamefont{{Yungelson}}},
  \bibnamefont{and} \bibinfo{author}{\bibfnamefont{S.~F.}
  \bibnamefont{{Portegies Zwart}}}, \bibinfo{journal}{A\&A}
  \textbf{\bibinfo{volume}{375}}, \bibinfo{pages}{890} (\bibinfo{year}{2001}),
  \eprint{arXiv:astro-ph/0105221}.

\bibitem[{\citenamefont{O'Shaughnessy et~al.}(2007)\citenamefont{O'Shaughnessy,
  Kalogera, and Belczynski}}]{OshKalBel2007}
\bibinfo{author}{\bibfnamefont{R.}~\bibnamefont{O'Shaughnessy}},
  \bibinfo{author}{\bibfnamefont{V.}~\bibnamefont{Kalogera}}, \bibnamefont{and}
  \bibinfo{author}{\bibfnamefont{K.}~\bibnamefont{Belczynski}},
  \bibinfo{journal}{The Astrophysical Journal} \textbf{\bibinfo{volume}{667}},
  \bibinfo{pages}{1048} (\bibinfo{year}{2007}).

\end{thebibliography}

\end{document}